\documentclass[12pt]{article}
\usepackage{fullpage}
 \usepackage{url}
 \usepackage{graphicx}
\usepackage{lineno}
\usepackage{epstopdf}
\usepackage{caption}
\usepackage{natbib}
\newcommand{\MnervFull}{\emph{Megalonaias nervosa}}
\newcommand{\Mnerv}{\emph{M. nervosa}}
\newcommand{\Unio}{Unionidae}
\newcommand{\Rebekah}{Rebek\mbox{}ah }
\newcommand{\Cgig}{\emph{C. gigas}}

\newcommand{\Pstreck}{\emph{Potamilus streckersonii}}

\usepackage[ocgcolorlinks,citecolor=blue]{hyperref}
\usepackage{lineno}
 \usepackage{setspace}
       
\begin{document}
\author{\small\Rebekah L. Rogers$^{1*}$, Stephanie L. Grizzard$^2$, and Jeffrey T Garner$^{3}$}

\title{Strong, recent selective sweeps reshape genetic diversity in freshwater bivalve \MnervFull}
\date{}

\maketitle
\noindent1.  Dept of Bioinformatics and Genomics, University of North Carolina, Charlotte, NC 28223\\
\noindent2.  Department of Biological Sciences, Old Dominion University, Norfolk, VA \\
\noindent3. Division of Wildlife and Freshwater Fisheries, Alabama Department of Conservation and Natural Resources, Florence, AL \\

\noindent*  Corresponding Author: Rebekah.Rogers@uncc.edu \\


\noindent Short Title: Selective sweeps in \MnervFull \\

\clearpage

\subsubsection*{Abstract}
Freshwater Unionid bivalves have recently faced ecological upheaval through pollution, barriers to dispersal, harvesting, and changes in fish-host prevalence. Currently, over 70\% of species in North America are threatened, endangered or extinct.  To characterize the genetic response to recent selective pressures, we collected population genetic data for one successful bivalve species, \MnervFull.  We identify megabase sized regions that are nearly monomorphic across the population, signals of strong, recent selection reshaping diversity across 73Mb total.  These signatures of selection are greater than is commonly seen in population genetic models.  We observe 102 duplicate genes with high dN/dS on terminal branches among regions with sweeps, suggesting that gene duplication is a causative mechanism of recent adaptation in \Mnerv.  Genes in sweeps reflect functional classes important for Unionid survival, including anticoagulation genes important for fish host parasitization, detox genes, mitochondria management, and shell formation.  We identify sweeps in regions with no known functional impacts, suggesting mechanisms of adaptation that deserve greater attention in future work on species survival.  In contrast, polymorphic transposable elements appear to be detrimental and underrepresented among regions with sweeps.  TE site frequency spectra are skewed toward singleton variants, and TEs among regions with sweeps are present at low frequency.  Our work suggests that duplicate genes are an essential source of genetic novelty that has helped this successful species succeed in environments where others have struggled.  These results suggest that gene duplications deserve greater attention in non-model population genomics, especially in species that have recently faced sudden environmental challenges.

\clearpage

\subsection*{Introduction}

The origins of genetic innovation during ecological change remain among the most challenging questions in evolutionary theory.   The ways that genetic variation appears in populations and then responds to strong shifts in selective pressures is fundamental to understanding how organisms evolve in nature \citep{ranz2012}.  Among classes of mutations that can contribute to adaptation, duplicate genes and related chimeric constructs are held as a key source of innovation \citep{Ohno1970,conant2008,Rogers2017Exp,Rogers2011}.  

Theory suggests that gene duplications can create redundancy that frees sequences from selective constraint, allowing neofunctionalization and adaptive subfunctionalization \citep{Ohno1970,conant2008,des2008}. These mutations can produce novel proteins and novel changes in expression that are difficult to mimic via point mutations \citep{Rogers2017Exp,Rogers2011,Stewart2019}. Duplications may be less likely to be neutral than other classes of mutations, and may serve as mutations of large effect when large effects are needed \citep{Rogers2015Limit,schrider2010,Emerson2008}.     Transposable elements (TEs) and duplications show an interplay, where TE content and activity is expected to produce higher duplication rates \citep{bennetzen2000,yang2008}.  How these mutations contribute to evolutionary outcomes as selective pressures are altered is key to understanding the role that genetic novelty plays in adaptation.   Characterizing the genetic response to strong selection will clarify where the limits may lie in survival under fundamental shifts in environment. 

The final outcomes of strong selection during habitat changes can be observed in the spectrum of diversity after selection \citep{tajima1989,Nielsen2005,hartl2020primer}.   Regions of the genome that have contributed to adaptation will show reductions in nucleotide diversity across linked regions and highly skewed site frequency spectra \citep{tajima1989}.   Genome scans with population genetics can clarify what genetic variation has contributed to adaptation in a way that is agnostic to function and free from human-centric biases regarding animal survival \citep{Nielsen2005, ellegren2014}.  Examining genetic diversity among regions targeted by selection, we can discern the types of mutations and functional classes that are most important for survival and reproduction \citep{Nielsen2005,ellegren2014}.  Such approaches are commonly implemented in population genetic model systems, where sequencing and functional analysis are straightforward \citep{Sella2009,Rogers2010,Rogers2011, DPGP,DGRP}.   

As genome sequencing has advanced, similar analysis can identify outcomes of selection in new organisms \citep{ellegren2014} where extreme environmental shifts have resulted in challenges to species survival.  These technological advances allow us to finally explore questions of genetic novelty in alternative evolutionary systems with species that have faced sudden shifts in selective pressures.  These goals will help clarify whether the genetic response during adaptation is fundamentally different under ecological upheaval compared with species that experience ecological stability.  The evolutionary response to ecological changes is often studied under deep time using phylogenetics.   However, in the modern anthropogenic era, we can observe challenges to species under threat happening in a single human lifetime \citep{Lake2000}.  This unfortunate opportunity allows us to study how species respond over shorter timescales than before, and determine genetic factors that allow some organisms to adapt while other species go extinct. 

\subsubsection*{Unionidae as a population genetic model}
Unionidae are notable for their unusual life cycle.  Adults are benthic filter feeders that settle into the substrate \citep{williams2008}.  Most species are dioecious with separate male and female sexes \citep{williams2008}.  Females are fertilized internally and brood offspring in their gills until releasing them into the water column where they parasitize fish hosts \citep{williams2008}.  As larvae mature, they drop off fish hosts to develop into mature adults \citep{williams2008}.  Phenotypic studies reveal arms races with fish hosts.  Unionidae are also unusual for their dual uniparental inheritance of mitochondria, where males inherit mitochondria from the paternal lineage and females inherit mitochondria from the maternal line \citep{Liu1996, Wen2017,Breton2007}.  Karyotype data for Unionida show no evidence of heterogametic sex chromosomes \citep{kongim2015}.  These unusual biological processes offer a unique setting to study parasite-host co-evolution, parental care, a transition from filter feeding to blood feeding, and unusual mitochondrial function.  

Freshwater bivalves, Unionidae, represent one clade that has recently experienced ecological upheaval.  Over 70\% of species in North America are threatened, endangered or extinct \citep{williams1993,Strayer2004,Haag2014,regnier2009}.  Other species have thrived even in the face of these same ecological pressures, some even experiencing recent population expansion.  SNP chips \citep{Pfeiffer2019}, single locus studies \citep{Pfeiffer2018}, mtDNA \citep{Campbell2005}, and reference genomes for the clade are just now being initiated \citep{Renaut2018,Rogers2021,Smith2021}, opening doors for evolutionary analysis in Unionid bivalves. These resources can clarify phylogenetic relationships, especially in cases where morphology, mtDNA and nuclear loci might differ \citep{Pfeiffer2019,Campbell2005}.   However, to gather a complete portrait of genetic novelty and the adaptive response to strong shifts in selection, whole genome population samples are required.

Muscle Shoals offers a focal location, with high species diversity for freshwater bivalves.  Historically over 80 species have been identified in the Tennessee River and surrounding tributaries \citep{williams2008}.   Rivers in the area have been dammed, preventing dispersal of aquatic fauna with impacts on species diversity \citep{Ortmann1924, Cahn1936}.  Water quality has been affected with pesticide and fertilizer runoff and industrial pollution, threatening avian and aquatic wildlife \citep{woodside2004}.   Bivalves experienced additional pressures, as they were harvested for shells in the freshwater pearl and button industries, with up to 6,700 tons produced per year during peak historical demand \citep{williams2008}.   Many freshwater bivalves in the Southeast now compete with invasive species such as \emph{Corbicula fluminea} (Asian clams) \citep{Strayer1999} and \emph{Dreissena polymorpha} (Zebra mussels) that have just recently been observed in the Tennessee River.  Some 32 species have not been seen at Muscle Shoals since the 1930s, with another 10 species listed as federally endangered \citep{Garner2001}.  Current estimates suggest roughly 40 species remain \citep{Garner2001}.  

During these modifications, other species that were able to thrive in flooded overbank habitats experienced intense population expansion \citep{williams2008,Ahlstedt1992,Garner2001}. Among these, \Mnerv {} is a success story where populations have expanded instead of contracting.   Surveys suggest that \Mnerv {} is among the most common species with tens of millions of individuals in one reservoir \citep{Ahlstedt1992}.  As a benchmark for adaptation in this clade where so many species are under threat, we have assembled and scaffolded a reference genome \citep{Rogers2021} and produced a whole genome population sequencing panel for \Mnerv.    Using scans of selection, we can identify adaptive genetic variation that has recently spread through populations due to natural selection \citep{Nielsen2005,ellegren2014}.  

In the data presented here, we observe signals of strong, recent selective sweeps reshaping genetic diversity in freshwater bivalves that have recently experienced ecological upheaval.  We use a genes-up approach that is agnostic to function, yet still observe genetic variation that recapitulates the biology of the organism.  In addition we identify targets of selection that have no clear functional annotation.  These \emph{loci} reveal as yet poorly understood mechanisms of adaptation in bivalves that deserve future study to discover the full genetic response to strong, recent selection. We observe key contributions from duplicate genes, which serve as causative agents of adaptive changes in \Mnerv, including functional classes known to influence the biology of the organism.  Together, these results indicate that duplicate genes are an essential part of the evolutionary response to strong selection that deserve greater attention in conservation genomics and non-model evolutionary genetics.

\subsection*{Results}
Across the genome of \Mnerv {} Tajima's $D$ is negative on average with a mean of -1.16 and a median of -1.2 suggesting population expansion in the past.   Mean diversity $\pi$ is $\theta=0.0054$ (Table \ref{Pgen}), slightly lower from single genome estimates of heterozygosity in a deeply sequenced reference specimen \citep{Rogers2021}.  Simulations of a 10-fold population expansion roughly 5000 generations ago recapitulate background diversity measures (FIgure \ref{PopGrowth5000}, Supplementary Information).   Inference of ancient population expansion is consistent with a species range expansion at the end of the last glaciation as greater habitat became available after glaciers receded \citep{keogh2021,elderkin2007}.  Diversity $\pi$ and Tajima's $D$ varies across scaffolds, with a roughly normal distribution genome wide (Figure \ref{normal}).  We used the heterozygosity at 4-fold non-synonymous sites where any of the 3 possible nucleotide substitutions alters amino acid sequence compared with 4-fold synonymous sites where none of the nucleotide alters amino acid sequence to measure constraint in the \Mnerv {} genome.  We find that $\pi_N/\pi_S = 0.47$ and $S_N/S_S = 0.48$.   These numbers are consistent with estimates from the reference genome suggesting  $H_N/H_S = 0.46$.  

We identified the Site Frequency Spectrum (SFS) for putatively neutral 4-fold synonymous sites and for fourfold nonsynonymous sites.  The SFS suggests a modest difference in the impacts of amino acid changing SNPs compared with synonymous SNPs in its slightly more U-shaped spectra with greater numbers of highest and/or lowest frequency alleles, but differences are not significant (Wilcoxon Rank Sum Test $W = 1331540915$, $P = 0.1086$; Kolmogorov-Smirnov $D = 0.0066259$, $P = 0.2524$; Figure \ref{SFS}). Such results indicate mild constraint on polymorphism across the genome. 

\subsubsection*{Strong, recent selective sweeps}

\begin{figure}
\begin{center}
\includegraphics[scale=.7]{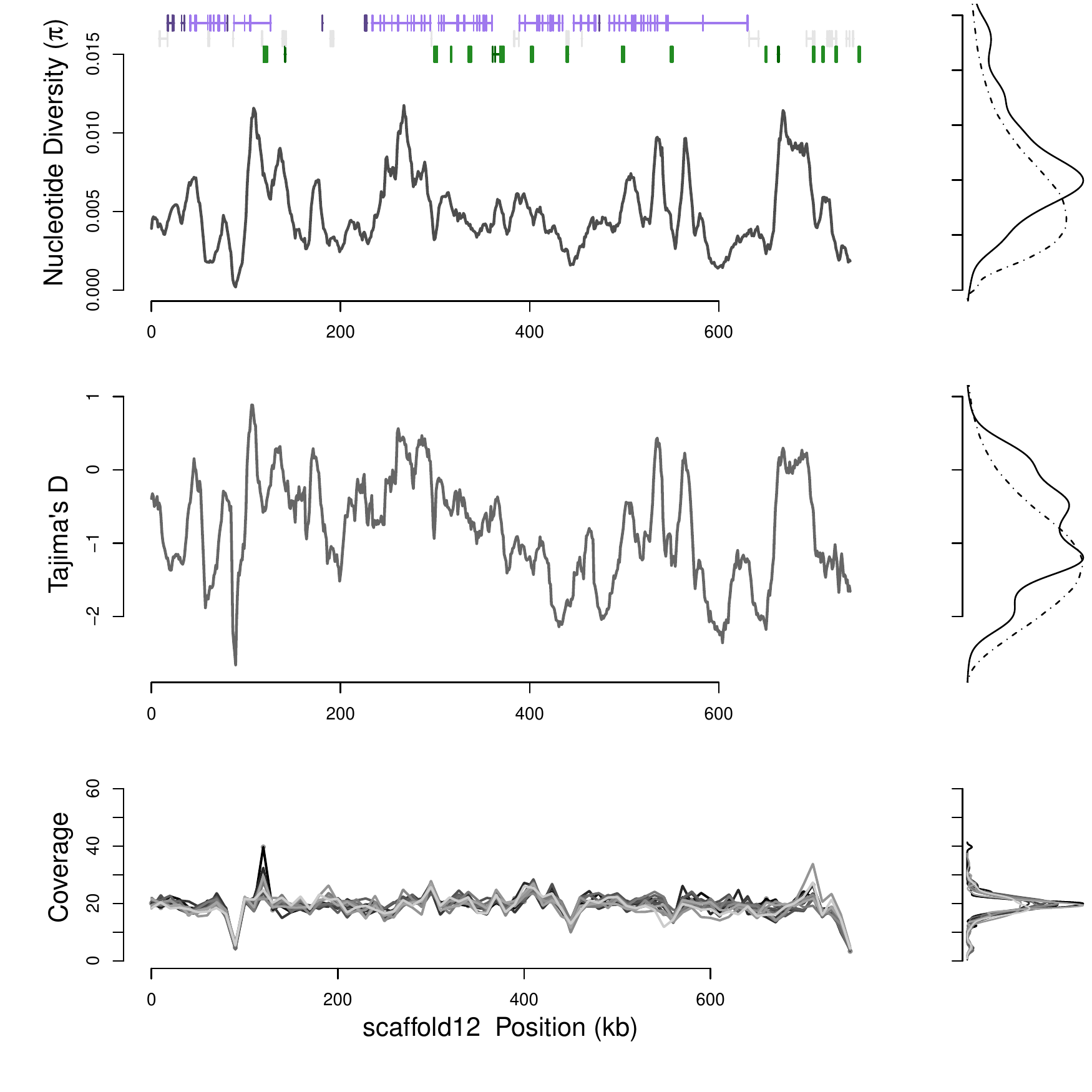}
\end{center}
\caption{Population genetic diversity $\pi$, Tajima's $D$, and normalized sample coverage for for a typical scaffold in \Mnerv.  Density plots of each metric are shown to the side for the scaffold (solid line) and genome wide background (dashed line).  Genetic diversity, $\pi$ is centered about 0.0054 and varies across the genetic region.  Normalized coverage for the 13 samples (shaded grey lines) shows a single polymorphic CNV/repeat around 470kb. Gene models confirmed with RNAseq data are shown in purple, unconfirmed in RNAseq data shown in grey, and repetitive elements in green.    \label{normal}}
\end{figure}

\begin{figure}
\begin{center}
\includegraphics[scale=1]{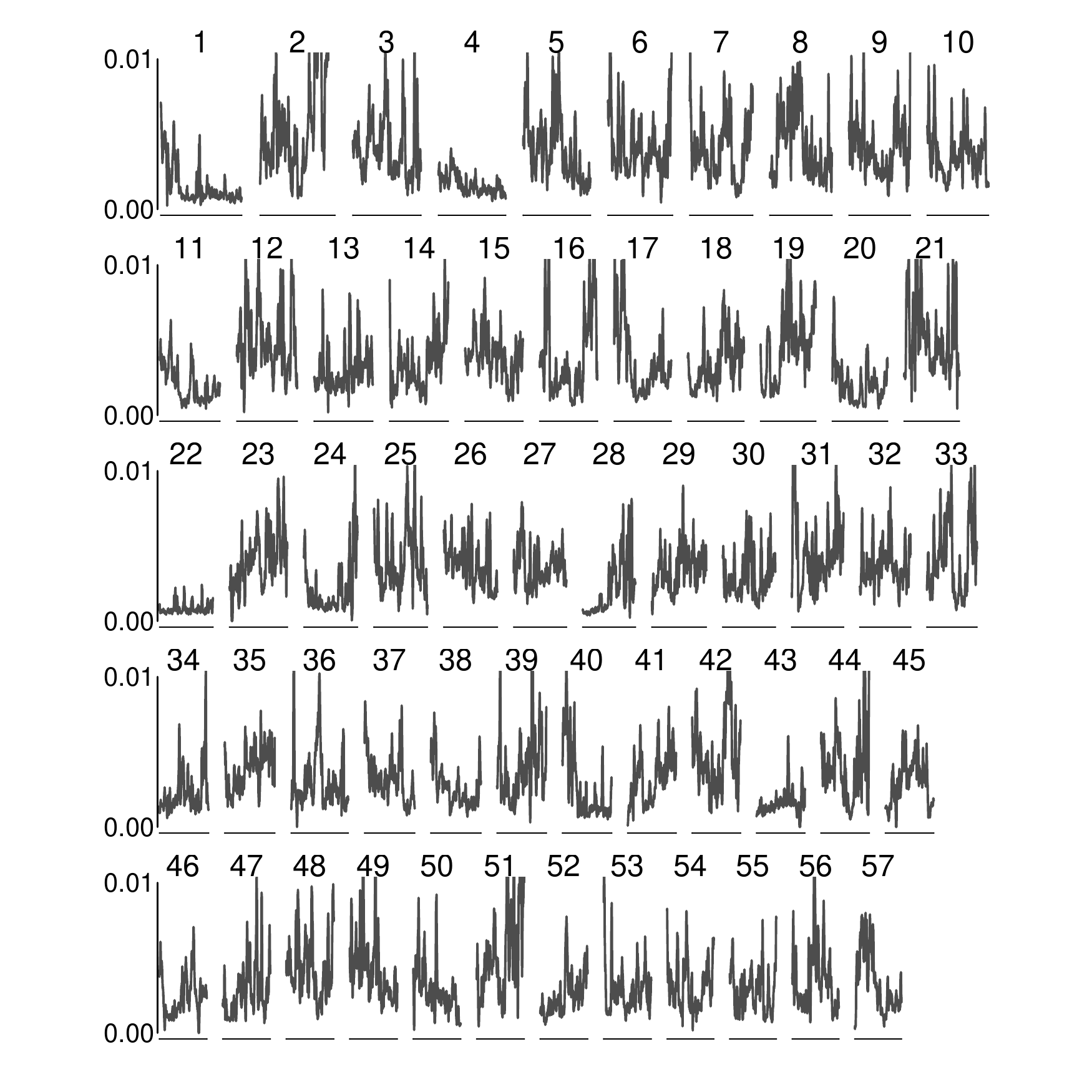}
\end{center}
\caption{Nucleotide diversity $\pi$ on the 57 largest genome scaffolds in order of scaffold size. Scaffold 4, Scaffold 22, Scaffold 43, and most of Scaffold 20 are nearly monomorphic across most of the sequence.  Scaffold 28 begins with reduced diversity and then returns to normal levels of polymorphism.  Diversity varies between 0.00 and 0.01, with mean genetic diversity of 0.0054. The largest scaffold shown is 980 kb, and the shortest is 579 kb. \label{brady}}
\end{figure}

We used nucleotide diversity $\pi$, $\theta_W$ and Tajima's $D$ to identify hard selective sweeps with signatures of natural selection altering genetic diversity in the population of \Mnerv.  We identify signatures of very strong, very recent selective sweeps in \Mnerv {} that have reshaped diversity in \Mnerv {} (Figure \ref{brady}).  Across the \Mnerv {} genome 73 Mb on 851 contigs is contained in regions that are borderline monomorphic for extended tracts (where $\pi < 0.0015$).   These regions constitute 2.7\% of total genome (2.6 Gb) and 6.2\% of assayable sequence in scaffolds 10 kb or longer (1.16 Gb).  Such signals of genetic diversity show strong linkage at swept regions, and are in addition to putative ancient selective sweeps (Figure \ref{normal}).  Regions with sweeps include 31.5\% coding sequence compared with background rates of 35.7\%.   Results are not explained by deletions, low coverage or unidentified repetitive sequence (Figure \ref{monomorphic}). The presence of \Cgig {} orthologs and typical genomic coverage confirms that these sequences are in fact of mollusc origin, not driven by contaminants. Specimens are different ages (Table \ref{Specimens}) and fish host dispersal \citep{williams2008} or even avian transport \citep{Darwin1878} minimizes the likelihood of incidental relatedness from microgeographic effects.  The genus shows little geographic differentiation \citep{Pfeiffer2018}. These extended tracts of low genetic diversity are unusual in comparison with scans of selection in standard population genetic models \citep{DPGP, DGRP, Rogers2010, Rogers2011,Sella2009}. 

Historical data are well documented, and are not consistent with population bottlenecks \citep{isom1973,Garner2001}.  However, we performed simulations that suggest even an unobserved 5-fold reduction in population sizes cannot explain these sweep like signals in the population (Figure \ref{ModBottleneck}, Supplementary Information).  Ancient bottlenecks do not reduce genetic diversity to the levels close to zero, as observed in sweep-like signals (Figure \ref{ModBottleneck}, Supplementary Information).  Recent bottlenecks are expected to reduce population diversity according to $H_t=H_0 (1-1/2N)^t$.  Assuming 20 generations since the formation of the dams along the Tennessee River \citep{Ortmann1924} even a 10 fold reduction in population size would not reduce diversity below 99.96\% of normal levels in the past century.  This effect is less than the rounding error diversity estimates.  Coalescent times in \Mnerv {} are expected to be 777,000 generations \citep{Rogers2021}, suggesting that neutral genetic diversity captures demographic effects far in the past prior to formation of dams.  Hence, we suggest that neutral forces of genetic drift act too slowly to reshape genetic diversity in this species on historical timescales.  

In contrast, very strong selective sweeps can produce 1-2Mb reductions in diversity even in less than 20 generations, though sweeps further in the past show similar patterns (Supplementary Information).  Hence we suggest that these reductions in diversity indicate a response to strong selective pressures, in recent evolutionary time.  While we do not have sufficient resolution to differentiate between sweeps on 10 generations compared to 1000 generations, the results are compatible with genetic changes occurring on historical timescales or in the recent evolutionary past.    We would suggest that these genomic regions contain alleles that were essential to survival or reproduction in the recent past.   


The \Mnerv {} reference genome has improved with additional long read sequencing (Supplementary Information), but still has an N50 of 125 kb. A total of 96\% of the genome is in scaffolds 10 kb or larger, sufficient to estimate population genetic statistics and 80\% is now in scaffolds 50 kb or larger. To help anchor sweep regions, we used syntenic comparisons with the more contiguous genome of \Pstreck {} (N50=2Mb) \citep{Smith2021} to align scaffolds likely to belong in the same genomic regions.   One selective sweep identified spans the majority of five scaffolds, with 1.6 out of 1.9Mb nearly monomorphic across the region (Figure \ref{monomorphic}). This scaffold (Scaffold 22) contains \emph{Furin},  upstream in a pathway from von Willebrand factors that are known to have rampant duplication and elevated amino acid substitutions across paralogs  ($dN/dS>1.0$) \citep{Rogers2021}.  

We identify 851 regions in extreme selective sweeps. Only 350 of these have a gene annotation in them with functional info in \Cgig {} \citep{oyster2012} or from Interproscan.  Sweeps span 0-9 genes.  Some 185 regions have only 1 functional annotation based on comparison with \Cgig {} \citep{oyster2012} or Interproscan annotations. Homology mediated annotation may be more limited for this species, as well annotated marine relatives are more than 200 million years divergent \citep{bolotov2017,TimeTree}.   For these regions containing only a single gene, we would suggest that it is most likely the causative agent of the selective sweeps.  Examples of two such single-gene sweeps include a chitin synthtase  (Figure \ref{chitinsweep}) and a cytochrome P450 gene (Figure \ref{cytsweep}).  Both represent categories that have experienced adaptive gene family amplification \citep{Rogers2021}.   Functional categories in regions with strong, recent selection include functions involved in shell formation, toxin resistance, mitochondria, and parasite-host co-interactions like molecular mimicry or anticoagulation, and stress tolerance (Table \ref{GenesInSweeps}).  These classifications are consistent with adaptive gene duplications in the reference genome \citep{Rogers2021}.  Curiously, ABC transporters do not appear multiple times in sweeps, in spite of rampant duplication and their known interactions with cytochromes.  

Genes in sweeps also include categories of DNA repair, development, apoptosis inhibitors (Table \ref{GenesInSweeps}), as well as less well characterized functional categories like Zinc fingers or Zinc knuckles, cell adhesion, cytoskeleton proteins, and WD-40 repeats.  While these conserved domains have less clear functional impacts based on the current information, but point to putative coevolution of large protein complexes, the exact nature of which remains unknown.   Other regions have no known conserved domains or functional annotation, suggesting they contribute to bivalve survival, even when we cannot explain why. Duplicate gene analysis and selective sweeps were performed in a high throughput, whole genome setting without preconceived bias regarding which functional classes should be represented.  Yet, in these scans of selection we identify functional classes that reflect the biology and environmental history of the organism. Here, these computational approaches can offer a more complete account of factors that are important for organism survival and reproduction. As Unionid genomics improves and greater functional information is provided, we may be able to resolve what other biological functions have contributed to adaptation in Unionidae.


%

\subsubsection*{Adaptive Gene Duplication}
\begin{figure}
\begin{center}
\includegraphics[scale=.7]{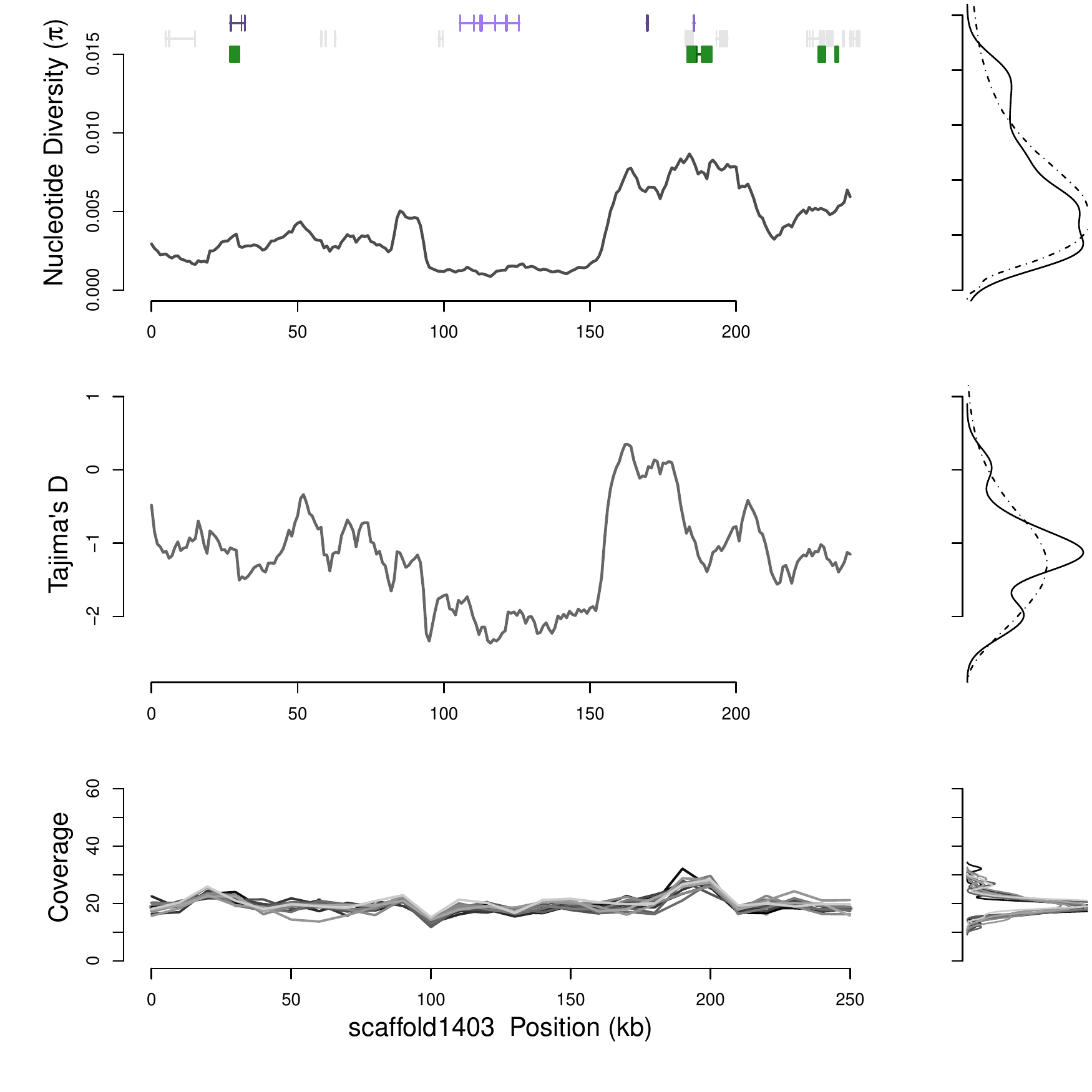}
\end{center}
\caption{Population genetic diversity $\pi$, Tajima's $D$, and normalized sample coverage for a chitin synthetase gene.   Density plots of each metric are shown to the side for the scaffold (solid line) and genomewide background (dashed line). Normalized coverage for the 13 samples (shaded grey lines) shows a small polymorphic deletion at 100 kb and a small polymorphic duplication at 190 kb.  The chitin synthetase is the only functional annotation within this region. \label{chitinsweep}}
\end{figure}

\begin{figure}
\begin{center}
\includegraphics[scale=.7]{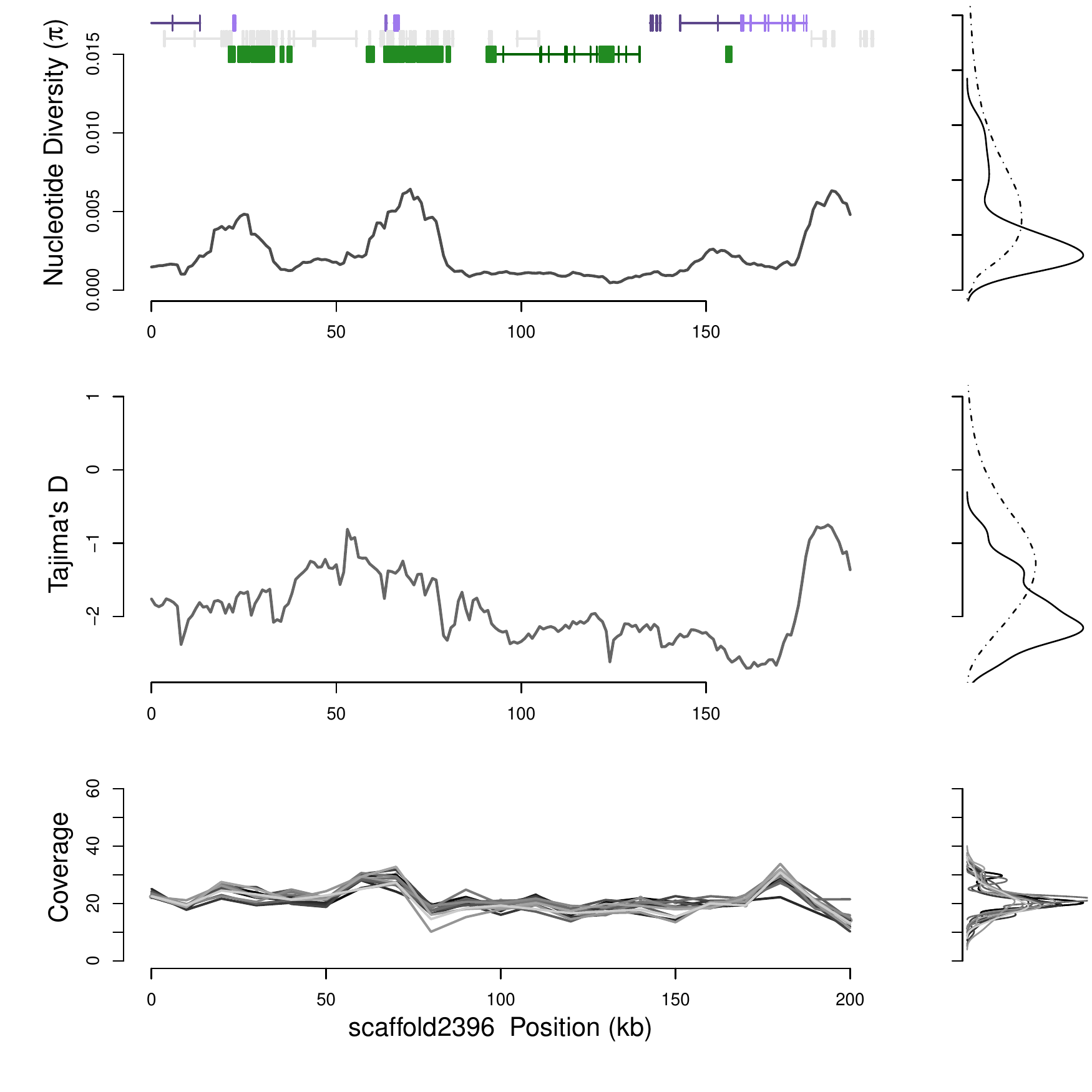}
\end{center}
\caption{Population genetic diversity $\pi$, Tajima's $D$, and normalized sample coverage for a selective sweep at the locus of a Cytochrome P450 gene in \Mnerv.  Density plots of each metric are shown to the side for the scaffold (solid line) and genomewide background (dashed line). Such signals would require strong, recent selection to substantially alter genetic diversity for an extended region.  The cytochrome gene is the only functional annotation within this selective sweep. Normalized coverage for the 13 samples (shaded grey lines) shows small polymorphic CNV/repeats at the locus of a TE around 60kb that drives a brief spike in diversity.  \label{cytsweep}}
\end{figure}

We identified gene families in re-annotated sequences in version 2 of the \Mnerv {} reference genome and estimated dN/dS across paralogs (Supplementary Information), a signature of selection for adaptive bursts of amino acid substitutions \citep{Goldman1994}.  This metric of selection should offer  a gene-specific metric of selection.  We observe no correlation between dN/dS on terminal branches and nucleotide diversity $\pi$ ($P=0.89$, $R^2=-0.00047$), indicating that these two tests of selection are independent from selective sweep analysis (Figure \ref{PiDnDs}). These genes include Cytochrome P450 genes, von Willebrand proteins, and shell formation genes (Table \ref{GenesInSweeps}, Figure \ref{chitinsweep}-\ref{cytsweep}). We observe duplicate genes with high $dN/dS$ across paralogs that are also located in strong recent selective sweeps.   Hence, such genes are strong candidates of causative agents during adaptation as they have two independent signals of recent selection.  

We identify a total of 102 genes with signatures of selective sweeps and high dN/dS  on their own terminal branch. Terminal branches represent timescales most compatible with recent selective sweeps that can be assayed with population genetics.  Among these are 2 cytochromes, 1 von Willebrand factor, 1 chitin sythetase, and A-macroglobulin TED domain important for anticoagulation.  Others of unknown function include DUFs, WD40s, and Zinc Knuckles.  In addition, some sweeps contain adjacent copies of duplicate genes with the same functional annotation with no other causative factors, even when $dN/dS <1.0$.  These likely represent adaptation without the burst of amino acid substitutions. A total of 245 have $dN/dS>1.1$ on any past branch.  These include 2 additional von Willebrand, 2 more cytochromes, 1 Thioredoxin, and some DUFS, more Zinc Fingers, inhibitors of apoptosis, and mitochondrial genes.   

We identify these two independent measures of selection in elevated $dN/dS$ across paralogs and independently identified sweep signals from population diversity. Such concordance represents a rare case of multiple measures of selection pointing to  gene duplication as a causative source of recent adaptation in \Mnerv.  

\subsubsection*{TE insertions are detrimental}
\begin{figure}
\begin{center}
\includegraphics[scale=.75]{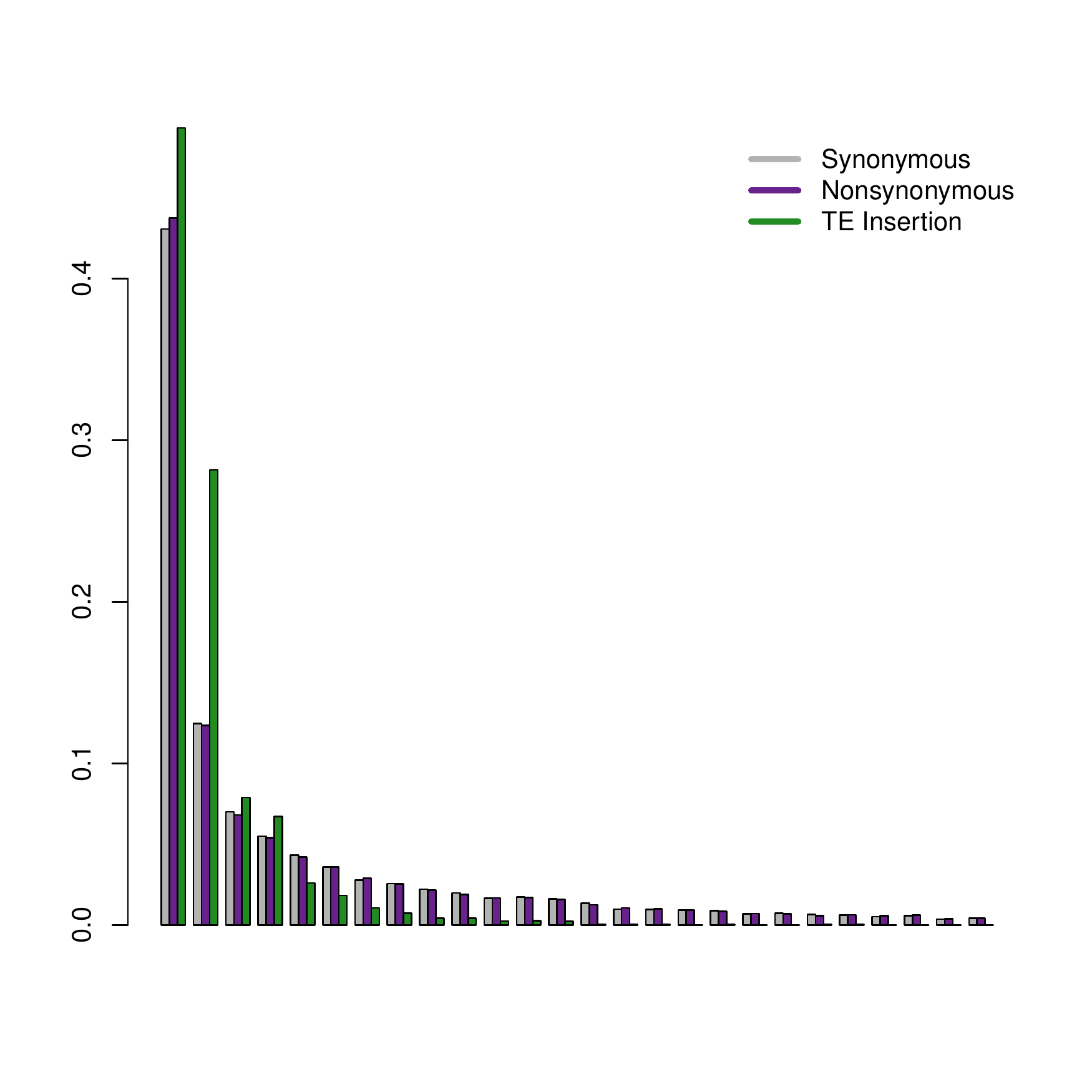}
\end{center}
\caption{\label{SFS} Site frequency spectrum for synonymous SNPs, nonsynonymous SNPs, and TE insertions in \Mnerv.  The SFS for TE insertions is skewed toward extreme allele frequencies, suggesting non-neutral impacts compared with synonymous and nonsynonymous SNPs.  Few polymorphic TE insertions reach moderate frequency and none are at high frequency in \Mnerv. }
\end{figure}

Transposable elements are selfish constructs that proliferate in genomes even at the expense of their hosts.  These repetitive sequences are intimately related to gene duplication rates, as they can facilitate ectopic recombination, form retrogenes, and translocate copies of neighboring DNA.  Signals of TE proliferation, especially of \emph{Gypsy} and \emph{Polinton} elements, were observed in the reference genome of \Mnerv {} \citep{Rogers2021}.  To determine whether these TE insertions may be adaptive or detrimental, we surveyed frequencies of TE insertions in the population of \Mnerv.   We identified genome rearrangements with abnormally mapping read pairs and used BLAST to identify those with transposon sequences at one of the two genomic locations.  We assume that the TE-associated region is the donor and the region without TE sequences is the acceptor region where the new TE copy lands.    We identify 4971 TE insertions that can be characterized in this reference genome.  TE insertions appear to be highly detrimental in \Mnerv, with a skewed SFS showing an excess of singletons that is significantly different from SNPs (Wilcoxon Rank Sum Test $W = 150176864$, $P < 2.2e-{16}$; Kolmogorov-Smirnov Test $D = 0.30443$, $P < 2.2e-{16}$ Figure \ref{SFS}). 

\begin{figure}
\begin{center}
\includegraphics[scale=.7]{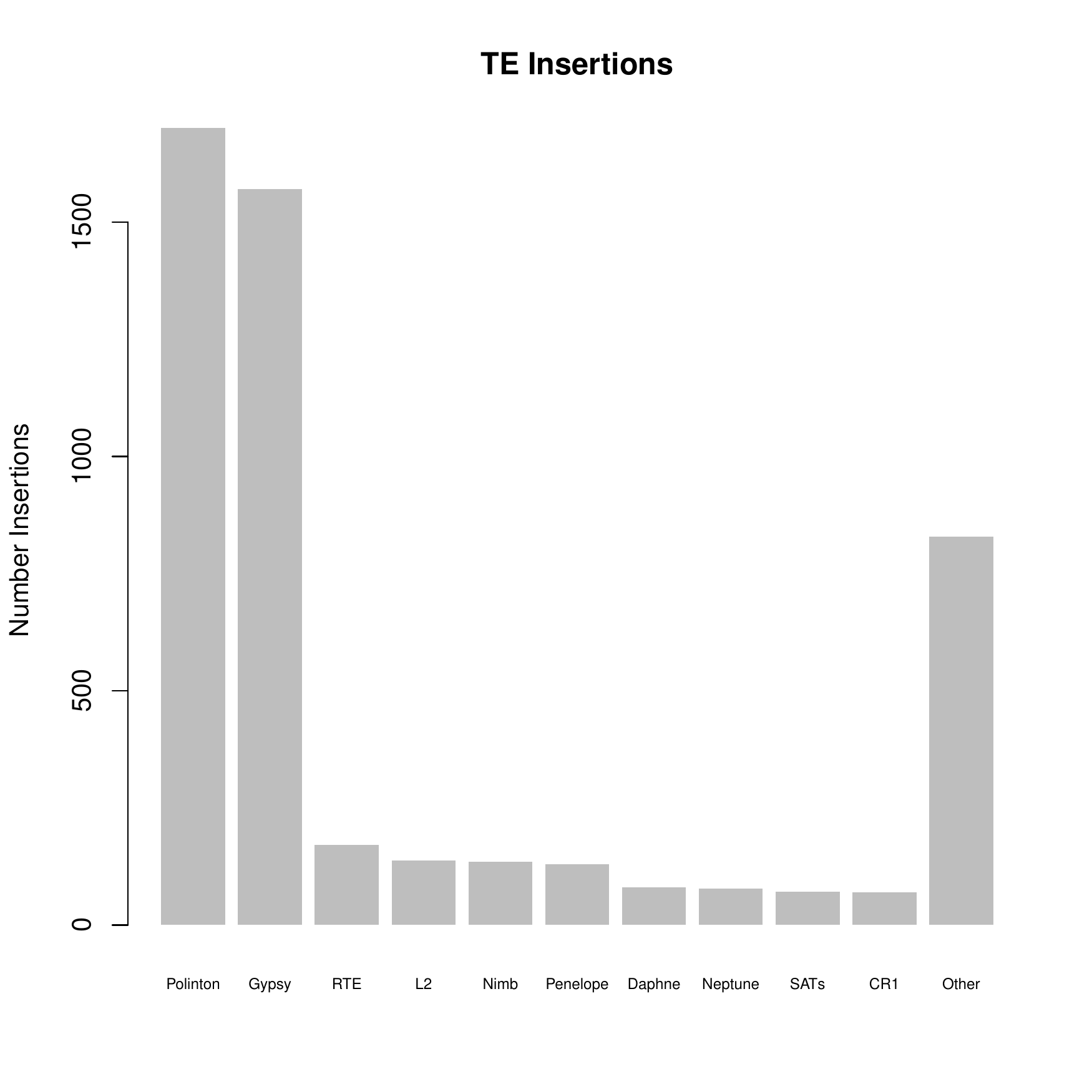}
\caption{\label{TETypeFig} Polymorphic transposable element insertions identified in populations of \Mnerv.  \emph{Gypsy} elements are most common, followed by \emph{Polinton} elements, consistent with recent TE proliferation identified in the reference genome of \Mnerv.  }
\end{center}
\end{figure}

TE movement is dominated by \emph{Polinton} DNA transposon activation and \emph{Gypsy} retroelements (Table \ref{TEType}, Figure \ref{TETypeFig}).    We also observe \emph{Neptune} element activity, consistent with the presence of these elements in the \Mnerv {} reference but not in other species surveyed \citep{Rogers2021}.  TEs are underrepresented among reference genome sequence in regions with selective sweeps ($P<10^{-5}$).  We find only 8.9 Mb (1.5\%) of TEs identified via RepDeNovo compared against an expectation of 34 Mb if TE content were allocated proportionally to the amount of the genome captured by sweeps.  We identify 18 Mb of RepeatScout TEs, against an expectation of 55 Mb TEs.   Only low frequency insertions at a frequency of 1/26-3/26 are identified in swept regions, precluding the possibility that these mutations are causative agents of selection in strong, recent selective sweeps.   Combined with the SFS, we conclude that these mutations are unlikely to be adaptive in \Mnerv, and are rather forming detrimental insertions throughout the genome.   

\subsection*{Discussion}
\subsubsection*{Selection under environmental upheaval}
Scans of selection can clarify genetic variation that has contributed to survival and reproduction, with fewer \emph{a priori} biases about what functions should be represented \citep{Nielsen2005,ellegren2014}.  This reverse ecological genetics can identify the most likely candidates driving selective sweeps, offering more complete information about how adaptation occurs in nature \citep{ellegren2014}.  Historical records suggest population expansion, excluding the possibility of a bottleneck event \citep{Garner2001,Ahlstedt1992}, an advantage for population genetics over model organisms that typically lack extensive historical or fossil records.  Moreover, simulations of bottleneck events do not show similar genetic signals compared with strong selective sweeps (Supplementary Information).  These selective sweeps occurred in recent evolutionary time, though resolution on historical time is difficult.  If historic DNA could be acquired for the same species at the same location, or through comparisons at other locations, it might resolve how much of these adaptive changes are the product of anthropogenic influence over the past 100 years.  

These results recapitulate and confirm the functional categories represented among adaptive gene family amplification in a single genome \citep{Rogers2021}. These selection scans offer greater detail and more information about adaptation than analyses that can be done with single reference genomes. We also identify strong selection in regions with less clear functional implications, such as WD-40 domains, Zinc Fingers, Zinc Knuckles, and apoptosis genes, or even regions with no known functional impacts.  These signatures of selection on regions with unknown functions that open questions remain that will need to be explored in future work to fully understand the drivers of genetic adaptation in Unionidae.   If future functional analysis can determine what lies in the regions with unannotated genes, we can better understand factors that contribute to success of \Mnerv {} and how to help endangered populations. 

\subsubsection*{Gene duplications and adaptation}
Gene duplications have long been held as a source of evolutionary innovation that can contribute new genes with novel functions \citep{Ohno1970,conant2008}.  Theory suggests that duplicate genes can produce new copies of genes that are functionally redundant.  Under reduced constraint copies may accumulate divergence and thereby explore novel functions.  Alternatively, duplicate genes may specialize in ancestral functions and offer adaptive subfunctionalization in escape from adaptive conflict \citep{des2008}.  Our results showing duplications followed by a burst of amino acid substitutions and strong signals of selective sweeps are consistent with these evolutionary models.  It has long been proposed that mutations of large effect appear first, that later are fine tuned through mutations with narrower functional impacts to reach evolutionary optima \citep{Orr2006}.  Empirical evidence suggests that such outcomes are a regular product of gene duplication in other systems \citep{jones2005, long1993, des2008,conant2008}.   Such models would be consistent with duplications serving as such mutations of large effect that later are fine-tuned through amino acid substitutions.   

It is striking to observe such independent signals of selection on multiple duplicate genes: high dN/dS across paralogs and presence in strong, recent selective sweeps.  Over 100 duplications contribute to recent adaptive changes with elevated dN/dS on the terminal branch, and 245 duplicate genes in recent selective sweeps are members of gene families with adaptive signals further in the past.  The genomic patterns are overwhelming that these mutations are key contributors to innovation in this species has experienced recent environmental threat. Duplication in gene families are important for the biology of \Mnerv {} appear to be  key for survival and reproduction in this robust species.  Shell formation genes, detox genes, and genes involved in parasite-host coevolution all are present in regions with selective sweeps.     Fish-host interactions, detox pathways and shell formation, suggest strong recent selective pressures consistent with known ecological challenges are reshaping genetic variation in \Mnerv.  Parallel analysis on marine bivalves has revealed adaptive duplications in adaptation to environmental changes \citep{sun2017, hu2022}.  Hence, we expect that these principles hold true outside this single species.  

We observe an association between detox genes and related stress resistance in a species that has recently been exposed to high levels of pesticides, herbicides, and other pollutants.  These chemicals gained widespread use at Muscle Shoals after the 1940s through mosquito and malaria control efforts , farming, and other industrial activities along the river \citep{woodside2004}.  The high pollution load has had extreme effects on wildlife throughout the region \citep{woodside2004}.  Pesticide and herbicide use is known to induce very strong selective pressures in other evolutionary systems.  Previous studies have observed parallel cases of detox gene duplication or rearrangements of lesser magnitude in \emph{Drosophila} \citep{schmidt2010,aminetzach2005}, Morning Glories \citep{van2020}, and rodents \citep{Nelson2004}. It is likely that the gene duplications observed in \Mnerv {} are an important part of the genetic response to these selective pressures.  

\subsubsection*{Transposable element bursts}
Transposable elements are selfish genes that amplify themselves even at the expense of host genomes.  They can break gene sequences, remodel expression for neighboring genes, and create chimeric TE-gene products \citep{schaack2010,Feschotte2008,dubin2018}.   It is hypothesized that TEs may offer a source of innovation, especially when species experience environmental stress.  Recent TE amplification for \emph{Gypsy} and \emph{Polinton} elements was previously observed in the reference genome of \Mnerv {} \citep{Rogers2021}, but it remained unclear whether such population expansion was adaptive or a detrimental byproduct of TE escape from silencing.  In this new population genetic data for \Mnerv, we observe no support for adaptive TE insertions.  

TE insertions appear to be detrimental with a skewed SFS and underrepresentation in selective sweeps.   A species with very large population sizes may be able to weed out detrimental TE insertions while retaining adaptive variation, especially if unlinked from beneficial variation.  In light of these results, it seems most likely that the recent proliferation of \emph{Gypsy} and \emph{Polinton} elements may have escaped conflict in the short term, but are prevented from spreading through populations under strong selective constraint.  

These two different classes of elements represent both LTR retroelement expansion and DNA transposon activity, rather than expansion only within a single class. \emph{Gypsy} elements carry chromodomains, that can modify heterochromatin organization and modify gene expression for neighboring genes, with more widespread impacts than local dynamics of gene damage \citep{Gause2001,Chen2001,Gao2008}. Under expectations of genetics arms races, we might expect strong selection to favor suppressors of TE activity in the future \citep{Cosby2019}. 

There may be interplay between TE content and duplication rates, as repetitive elements can facilitate gene family amplification \citep{bennetzen2000,yang2008}, but these are indirect effects not tied to individual transposable element copies.  The cost of genetic innovation and TE activity may be different in small populations where evolutionary dynamics may be more permissive to TE proliferation \citep{LynchBook}. Future analysis of the spectrum of TE variation in other species of Unionidae with different population dynamics will help answer questions about the interplay of TEs and adaptive duplication during habitat shifts.  Regardless, individual TE insertions in \Mnerv {} largely appear to be maladaptive. 

\subsubsection*{Implications for Imperiled Species}
 In scans of selection in \Mnerv {} we identify many genes that point to fish host interactions in glochidia, a known point of attrition for many species \citep{modesto2018}.   Strong selection at glochidia stages, may favor larger brood sizes to overcome attrition in some reproductive strategies \citep{Haag2013}.  \Mnerv {} females can produce 1 million offspring per reproductive cycle \citep{Haggerty2005}, a factor that may influence their success. Genomics also points to detox genes, shell formation genes, stress response genes, and mitochondria genes, all consistent with factors important for Unionidae biology.  

Not all species may have sufficient genetic variation to solve these challenges under shifting selective pressures, especially when population sizes are small \citep{Hermisson2005,Maynard1971}.  \Mnerv {} with its large population expansion represents a successful species that can be used as a genetic benchmark for adaptive changes.  In threatened or endangered species with smaller population sizes, it is possible that genetic variation may be limited, and genetic drift may impede adaptive walks.  In \Mnerv {} some gene duplication is likely to have been pre-adaptation prior to the modern era as $dS>0$.  The likelihood of adaptation from standing variation for gene duplications was high in \Mnerv {} \citep{Rogers2021}.  

As duplicate genes appear to be important sources of innovation, we suggest that they deserve better characterization across Unionidae and in other species experiencing ecological threat.  Species that lack gene duplications may struggle to survive in the face of these same environmental pressures.  Conversely, current results would suggest that new TE insertions are primarily rare and non-adaptive.  They may be a distraction from the genetic changes that are most essential for species survival.  Highly variable low frequency TEs like \emph{Gypsy} and \emph{Polinton} elements in \Mnerv {} may serve as poor markers for species tracking.  These insights and genetic resources can help conservation biologists working in mussel management as they design analyses to monitor and aid threatened or endangered species of Unionidae.  

Future cross species comparisons may help address the genetic basis of adaptation in this clade that has experienced recent environmental upheaval.  How do species with smaller population sizes respond under these environmental pressures?  What is missing from the genomes of species that are most threatened? If we can begin to approach these genetic questions for Unionidae, we can better understand how evolutionary processes differ in species that are under threat and determine how genomes influence species survival.   

\subsection*{Methods}
\subsubsection*{Specimen Collection}
JT Garner collected specimens of \MnervFull {} at Pickwick Reservoir (Table \ref{Specimens}).   The largest (Specimen \#3) was selected as the reference genome specimen, sequenced and annotated as previously described \citep{Rogers2021}.    Some 14 specimens were dissected and sequenced using Illumina short read sequences for population genomic data.  The reference specimen N50 in previous sequencing was roughly 50kb, adequate for many applications in evolutionary genomics, but with limited information regarding linkage across distance.  Scaffolding attempts with HiC and OmniC have not been successful, with zero cross-linked read pairs \citep{Rogers2021}. To improve the reference genome for population genetic applications, we generated additional long read sequence data to scaffold the assembly into longer contigs with an N50 of 120 kb.  We re-annotated according to previously used methods \citep{Rogers2021} similar to those used to reannotate non-model \emph{Drosophila} \citep{G3} (Supplementary Information). 

\subsubsection*{Identification of regions with extended selective sweeps}
Sequences were aligned to the reference genome, and used to estimate genetic diversity statistics commonly used population genetic inference $\pi$, Wattersons $\theta$, and Tajima's $D$. We required that windows have full coverage for at least 75\% of sites across all strains.  Population genetic statistics were corrected for the number of sites in 10kb windows with full coverage.  We used msprime v1.1.1  \citep{baumdicker2022} and SLiM v3.7 \citep{Messer2013,Haller2019,Haller2019tree} to model expectations of diversity under demographic scenarios and with natural selection for these populations. Additional detail is available in Supplementary Information.  We identified SNPs from 4-fold synonymous sites where none of the 3 possible nucleotide substitutions alter amino acid sequence, and 4-fold non-synonymous sites where any of the 3 possible substitution alters amino acid sequence and estimated site frequency spectra (SFS) for each and tested for significant differences using both Kolmogorov-Smirnov tests and Wilcoxon rank sum tests.   

Extended regions with reduced genetic diversity display stronger signals than the classic V-shape most often described for typical selective sweeps \citep{Sella2009, hartl2020primer}.  They do not fit with theoretical models that could be used to place boundaries on the timing and selection coefficient of selective sweeps using diversity and recombination rates \citep{kaplan1989,Sella2009}.  To objectively place boundaries on recent selective sweeps we use Bayesian changepoint statistics.  Changepoint statistics are agnostic to the direction, magnitude, and duration of effects.  They identify regions of the data that depart from the background patterns in the remainder of the data, where shorter signals with greater magnitude may be significant as can longer signals of lesser magnitude.   We identified changepoints and posterior means of $\pi$, and Tajima's $D$ for regions between changepoints in the R package \emph{bcp} \citep{BCP} (Supplementary Info).   Genome assembly for freshwater molluscs remains challenging because of repeats and difficulty of long molecule extraction, with limited N50s even using long read data \citep{Smith2021, Rogers2021, Renaut2018}. The best assembly to date is for \Pstreck, with an N50 of 2 Mb after 100X coverage of PacBio and 48X coverage of 10X sequencing \citep{Smith2021}. To anchor scaffolds identified in the longest sweeps we used syntenic mapping against this more contiguous bivalve assembly to identify sections of \Mnerv {} likely to be from similar genetic locations (Supplementary Information).

\subsubsection*{Adaptive gene duplications}
We identified 4758 gene families (with 2 or more paralogs) using a First-Order Fuzzy Reciprocal Best Hit Blast \citep{Han2009} on the re-annotated the scaffolded genome as per previous methods \citep{Rogers2021} (Supplementary Information).  Protein sequences were aligned in clustalw \citep{clustalw} then back-translated to the original nucleotide sequence.  Synonymous and non-synonymous substitutions were analyzed with the codeml package of PAML \citep{PAML} with the F1x4 codon model using the clustalw generated guide tree.   We excluded 44 out of 4758 gene families that proved computationally intractable, with failed alignments or failed PAML runs.   We then identified duplication events with elevated amino acid substitutions across at least one branch for paralogs, suggesting selection for amino acid replacements (high $dN/dS >>1.0$), a gene-specific measure of selection \citep{Goldman1994}.  The locations of these adaptive gene duplications were matched with locations of strong, recent selective sweeps to identify cases where they therefore likely contribute as causative agents of selective sweeps.

\subsubsection*{TE insertions}
We identified polymorphic TE insertions using a paired-end read approach \citep{cridland2013}. We identified polymorphic read pairs that map to different scaffolds, indicative of DNA moving from one location to the other.  We required at least 5 abnormally mapping read pairs support each mutation, clustering read-pairs within 325 bp, based on the Illumina sequencing insert size.  Mutations with insertion sites within 325 bp were clustered across samples.  We took 1000 bp on either side of each breakpoint and matched these in a tblastx against the RepBase database \citep{RepBase} at an E-value of $10^{-20}$, requiring hits at least 100 bp long to a repetitive element sequence on one side, but not both sides of rearrangements.  These mutations were considered to be novel TE insertions compared with the ancestral state.  We used coverage $>1.75X$ and $<5X$ whole genome background levels to identify homozygous mutations. 

 \subsubsection*{Acknowledgements}
We thank Cathy Moore for advice on molecular assays.   We thank Karen Lopez for support with local Nanopore sequencing at UNCC.  Jon Halter, Michael Moseley, Chad DeWitt, Chuck Price, and Chris Maher provided help with software installation and functionality on the UNCC HPC system. We thank the Duke University School of Medicine for the use of the Sequencing and Genomic Technologies Shared Resource, which provided Illumina sequencing services for \Mnerv {} genomes and transcriptomes.   De Novo Genomics generated a subset of the Oxford Nanopore sequences for this study.    All analyses were run on the UNCC High Performance Computing cluster, supported by UNC Charlotte and the Department of Bioinformatics. 

\subsubsection*{Data availability}
\MnervFull {} genomic sequence data are available in PRJNA646917 and transcriptome data are available at SRA PRJNA646778.  The \MnervFull {} genome assembly is available at PRJNA681519.  Additional sequences will be released upon acceptance for publication.  Data prior to peer review is available at: \url{https://www.dropbox.com/sh/mwenn3v57dfuyvm/AADKl-EQjH_7Ws0jn-MHaGXva?dl=0}.

\subsubsection*{Funding}
This work was supported by startup funding from the Department of  Bioinformatics and Genomics at the University of North Carolina, Charlotte. \Rebekah L. Rogers is funded in part by NIH NIGMS R35 GM133376. The funders had no role in study design, data collection and analysis, decision to publish, or preparation of the manuscript.    

\subsubsection*{Author contributions}
RLR designed experiments and analyses \\
JG collected specimens from wild populations \\
RLR and SLG performed experiments \\
RLR performed analyses with assistance from SLG\\
RLR and JG wrote and edited the manuscript with input from SLG \\

\bibliography{MnervPopGenRev1}
\bibliographystyle{apalike}

\clearpage

\clearpage

\clearpage

\clearpage
\begin{figure}
\begin{center}
\includegraphics[scale=.75]{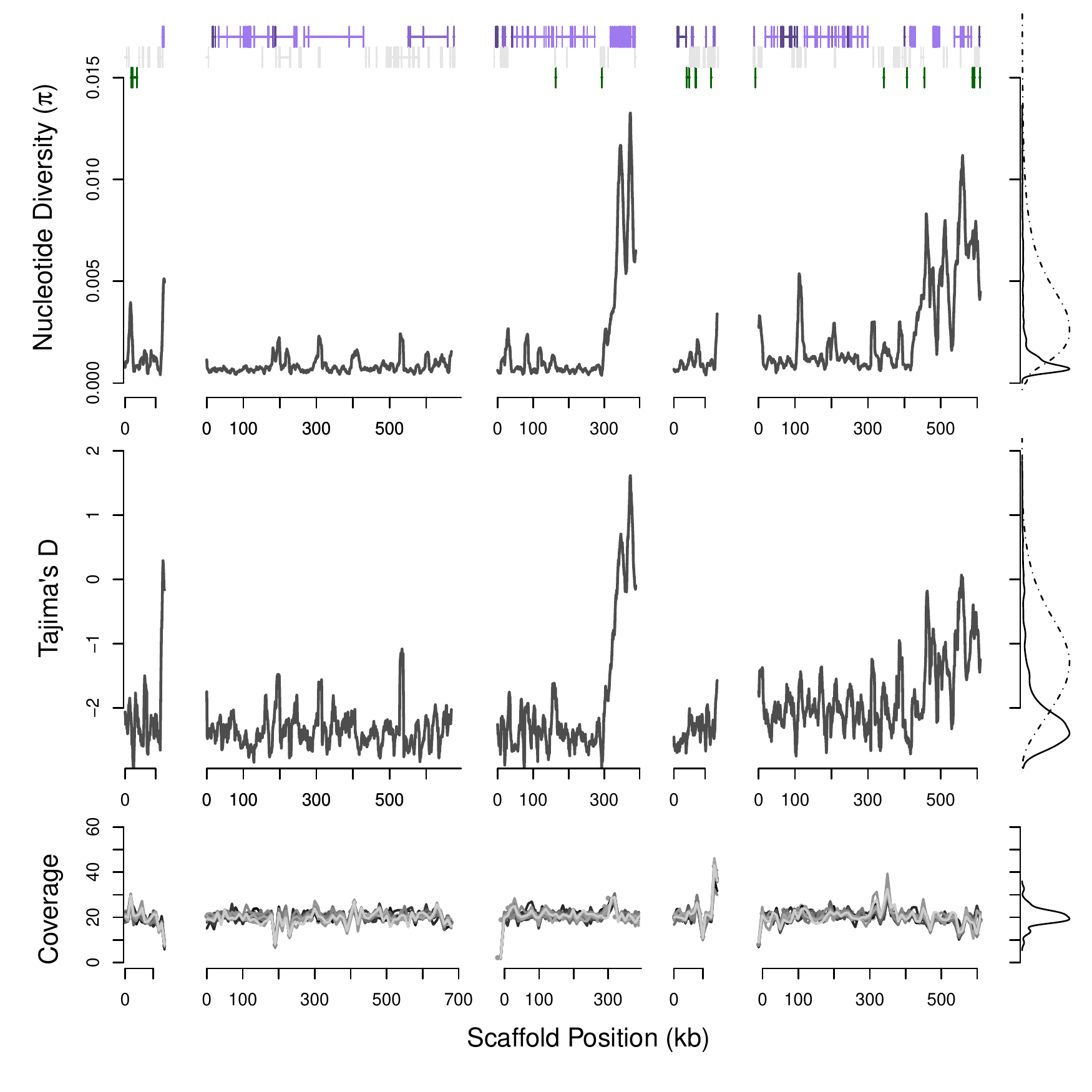}
\end{center}
\caption{Diversity on five scaffolds encompassing a 1.9 Mb region syntenic with \Pstreck {} is nearly monomorphic (scaffold4763, scaffold22, scaffold372 - reversed, scaffold4254, scaffold40-reversed).  Nucleotide diversity $\pi$ shows a marked departure compared to background genomic levels, as supported by changepoint statistics.  Tajima's $D$ is below -2.2 throughout the region.  Annotated genes match with marine bivalve \Cgig {}, eliminating possibility that results are driven by contaminating sequences.  The population genetic parameters suggest a strong, recent selective sweep that has driven out diversity from the population.  Genes in the region include anticoagulation gene \emph{Furin} on scaffold22. \label{monomorphic}}
\end{figure}

\clearpage

\clearpage

\clearpage

\clearpage{}

\renewcommand{\thefigure}{S\arabic{figure}}
\renewcommand{\thetable}{S\arabic{table}}
\setcounter{figure}{0}
\setcounter{table}{0}
\renewcommand{\thetable}{S\arabic{table}}
\renewcommand{\thepage}{S\arabic{page}}
\setcounter{page}{1}
\section*{Supplementary Information}
\subsubsection*{Specimen collection and processing}
JT Garner collected 20 female \MnervFull {} from Pickwick Reservoir near Muscle Shoals, AL.  Specimens were shipped in coolers to UNCC.  Four specimens died in shipping and two others had limited response to perturbation on arrival.  The other 14 were dissected and tissues were flash frozen in liquid nitrogen.   Length (anterior to posterior), depth (left to right shell, measured near the umbo), and height (dorsal to ventral) were measured with digital calipers at the maximum span of each shell. Ring counts were used to estimate age (Table \ref{Specimens}).  DNA was extracted from tissues using a Qiagen DNEasy column based purification kit.  Sequencing libraries were prepared in house using the TruSeq PCR-free library preparation kit and sequenced on 3 lanes of a HiSeq3000/4000 at Duke University Sequencing Core.  All specimens later thawed in a -80 meltdown.  Population specimens were intractable for RNA extraction and had very low yield for long molecule DNA extraction after thawing.  

\subsubsection*{Genome Scaffolding and Re-annotation}
A previously generated a reference genome is available with an N50 of 50 kb , sufficient to capture the majority of gene content in \Mnerv {} \citep{Rogers2021}. Linkage across the genome offers important information for population genetic models.  To improve this assembly, we performed 21 phenol-chloroform extractions on adductor tissue for the reference specimen that had thawed in a -80 meltdown.  We used bead-based cleanup to enrich for larger DNA fragments.  We performed long read sequencing with 1 full run and 3 incomplete runs of Nanopore on a MinIon. Basecalling was performed using the Guppy 3.6.1-cuda basecaller on GPU nodes with 4 cores.  Read sizes ranged from 200 bp to 329,099 kb.  

We scaffolded the prior assembly from \cite{Rogers2021} in LINKS/1.8.7, which identifies uniquely mapping k-mers to anchor scaffolds using long reads or long inserts.  Scaffolding was performed iteratively with subsets of reads for each run to manage memory usage. LINKS was run with distance thresholds of 2,000, 5,000, 10,000, and 20,000 bp using options -k 21 -t 5. This 2.6Gb improved reference assembly has an N50 of 125,303bp and maximum contig length of 992,038bp, far better performance for population genomics. 2.0 Gb out of 2.6 Gb of sequence on 15,643 scaffolds 50 kb or larger and 2.5Gb is on 33,101 scaffolds 10 kb or larger.  Hence, the majority of the genome is in contigs large enough for population genetic analysis.   BUSCO analysis v. 3.0.2 using the metazoan ortholog data set \citep{simao2015} shows 86.1\% complete single copy  orthologs, 2.0\% duplicated orthologs, and 5.6\% fragmented buscos.  This assembly has 50\% fewer fragmented orthologs than the previous assembly attempt \citep{Rogers2021}, but the number of orthologs found in whole or in part remains unchanged (only 6.3\% missing).  The number of duplicated orthologs remains unchanged.    Some sequences removed as putative contamination in the initial assembly now scaffold with mollusc sequences, suggesting prior filtering may have aggressively filtered out short repetitive elements that had incidental matches with contaminant genomes.  Liftover from the initial \Mnerv {} genome assembly, which contains identical contig sequences, was performed using a BLASTn \citep{BLAST}. The mtDNA scaffold scaffold21472.30567.f21430Z30567 was identified using a BLASTn \citep{BLAST} against \emph{Quadrula quadrula} mitochondria sequence obtained from NCBI (accession NC 013658.1, downloaded Aug 15 2018). 

We re-annotated sequences according to previous work \citep{Rogers2021}. We performed \emph{de novo} transcriptome assembly in Oyster River Protocol \citep{ORP}.  We mapped transcripts to the reference using BLAT \citep{kent2002} and used these as hints for Augustus allowing for isoform detection.  Augustus identified 263,584 putative gene sequences.  We used Interproscan to identify conserved functional domains \citep{quevillon2005}.  Transposable elements were identified using Interproscan \citep{quevillon2005}, tBLASTx \citep{BLAST} against RepeatScout TEs, RepBase \citep{RepBase}, and with RepDeNovo \citep{Chu2016}.  These methods identified 22,269 TE-related transcripts, with multiple transcripts identified for some TEs.  A total of 824 Mb of sequence matches with RepeatScout TEs in the reference genome, and 562 Mb matches with RepDeNovo repeats.  These were removed from putative gene annotations.  Orthologs and paralogs were identified in a BLASTp comparison at $E<10^{-20}$ against \Cgig {} \citep{oyster2012}. Genes with support from interproscan, blast comparisons with \Cgig, or from RNAseq data were kept in the final list of 64,086 putative genes.  64,086 genes encompassing 86,049 exons remain.   In this improved reference genome, we identify 172 Cytochrome P450 genes, 146 ABC transporters, 168 Mitochondria-eating genes, 55 thioredoxins, 109 Hsp70 genes, 282 rhodopsins, 92 chitin metabolism genes, and 242 von Willebrand genes.  This new assembly identifies even greater numbers of these highly proliferated gene families in \Mnerv {} than the previous version of the reference sequence even as the number of duplicated BUSCO orthologs decreased by 50\%.  Hence, we do not have any indication that duplicate gene content is decreasing as assembly quality improves.


A recent high quality reference assembly with greater contiguity (N50 of 2Mb) has recently become available for \Pstreck {} \citep{Smith2021} that can be used for comparison with other \Unio {} species. Using a reciprocal best hit BLASTn \citep{BLAST} at an E-value of $10^{-10}$, we identify 17862  1:1 orthologs between \Mnerv {} and \Pstreck.  We used these to create a syntenic map of scaffolds between these two species where sequence divergence is too great for nucleotide alignment. While rearrangements may have occurred during divergence of these two species, the syntenic map can be used to anchor scaffolds identified as subjects of strong selective sweeps.   

\subsubsection*{Genome alignment and Population Genetics}
We aligned DNA for 13 samples and the Reference specimen using bwa aln \citep{BWA}.  Sequencing depth was calculated using samtools depth.  We identified SNPs using gatk haplotype caller under default parameters in a diploid model.  We used SNPs to calculate $\pi$, $\theta_W$ \citep{watterson1975}, and Tajima's $D$ \citep{tajima1989} in 10 kb windows with a 1 kb slide.  Polymorphic TE insertions were identified using paired end read mapping and blast hits to RepBase as well as scaffolding with RepDeNovo contigs.  We estimated $\pi$, $\theta_W$ and Tajima's $D$ for each scaffold over 10kb in length using a window size of 10kb and a 1kb slide, excluding windows with less than 75\% coverage. 

Bayesian Changepoint statistics identify the most likely changepoints with posterior probabilities, and fits means to the blocks identified in the data using MCMC.    Because \emph{bcp} uses autocorellation to extract information about change points, only non-overlapping windows were used as input (10 kb regions with a 10 kb slide).  Chagepoints with greater than 75\% support were kept, and means of Tajima's $D$ and $\pi$.   Blocks within 20,000 bp were joined together to smooth stochastic switching.  We used a threshold of $\pi<0.0015$ for strong, recent selective sweeps, identifying 72 Mb of sequence show signals of extremely strong, extremely recent selective sweeps. Some 155 Mb on 1342 contigs is identified at a less stringent threshold of $\pi<0.0020$.

\subsubsection*{Population Genetic Simulations and Modeling}
We used msprime v1.1.1  \citep{baumdicker2022} and SLiM v3.7 \citep{Messer2013,Haller2019,Haller2019tree} to perform coalescent simulations and forward simulations of population genetic diversity.  We modeled genetic diversity $\pi$ and Tajima's $D$ under the following scenarios: 

\begin{enumerate}
\item population stasis
\item population expansion
\item bottlenecks
\item strong selective sweeps 
\item moderate selective sweeps
\item reduction in recombination
\end{enumerate}

\subsubsection*{Coalescent Simulations For Demographic Changes}
Using msprime v1.1.1 \citep{baumdicker2022} to perform coalescent simulations under different demographic changes.  Tajima's $D$ for \Mnerv {} is negative, with an average of -1.16.  Genome wide shifts in Tajima's $D$ toward negative values are a signal of population expansion.  Populations of \Mnerv {} are common throughout the eastern United states.  We simulate recent and ancient population expansion.  We used a baseline $N_e=300000$, consistent with estimates of population size from $\pi=0.0054$ and $\mu=5\times10^{-9}$.  Data were parsed in 10kb windows.  Coalescent and forward population genetic simulations require non-overlapping generations, which can be applied without severe loss of information \citep{baumdicker2022,Messer2013,Haller2019,Haller2019tree}.

\begin{figure}
\begin{center}
\includegraphics[scale=1]{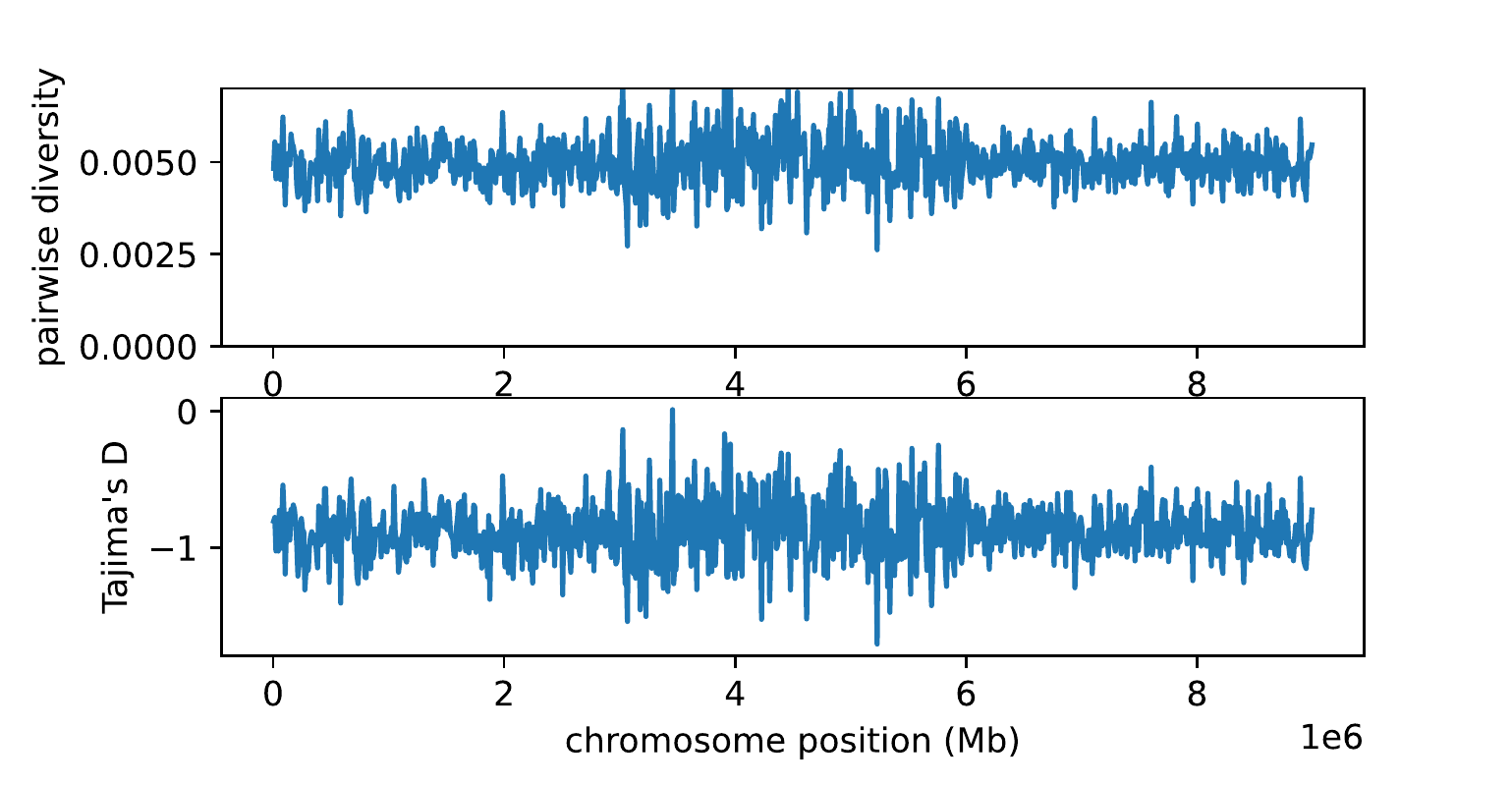}
\end{center}
\caption{Genetic diversity $\pi$ and Tajima's $D$ under simulated 12-fold population growth at a rate of 1.005 per generation over 5000 generations with a 5-fold population bottleneck for 10 generations,  20-10 generations in the past followed by population recovery. Simulations show highly recombining regions ($r=5\times10^{-8}$) from 0-3 Mb and 6-9 Mb, and lowly recombining regions ($r=5\times10^{-9}$) from 3-6 Mb.  \label{ModBottleneck}}
\end{figure}

A pattern of 10-fold recent population growth alone does not shift Tajima's $D$ sufficiently to match empirical observations (Figure \ref{ModBottleneck}).  Populations have insufficient time to accumulate new mutations that contribute to rare alleles after expansion.  However, a 12X population expansion beginning roughly 5000 generations ago can produce a shift in diversity $\pi$ consistent with empirical observations.  Both a sudden expansion at a rate of 25\% growth over 10 generations or prolonged population growth of 0.05\% per year over 5000 years can produce comparable shifts in Tajima's $D$ to approximately -1.0, with genetic diversity consistent with observations.  Population growth beginning 1000 generations in the past produces a lesser shift in Tajima's $D$ to 0.5.  At 10,000 generations, the shift in Tajima's $D$ is greater than observations in the data, at nearly -1.4.  

The timing of this ancient population expansion will depend on generation time.  \Mnerv {} has  a minimum age of reproduction at 4 years old, but typically reaches sexual maturity at age 10 \citep{woody1993}. This minimum of 4 years can be used to establish the absolute limits for shortest timescales of adaptation in recent time of 25 generations since the damming of the Tennessee River.  The maximum lifespan is 54 years, though 40 year old individuals may be more typical \citep{haag2012}.  Using the geometric mean of minimum sexual maturity and maximum lifespan, we obtain 15 years, but excluding outliers the geometric mean generation time would be 20 years. These simulations would be consistent with a range expansion in prehistoric time likely 75,000 years-100,000 years ago, depending on generation times of 15-20 years.   This ancient population expansion is consistent with a species range expansion at the end of the last glaciation as greater habitat became available after glaciers receded as seen in other species \citep{keogh2021,elderkin2007}. Differences in generation time might yield different scaling from time in generations in population genetic simulations to historical timescales. Pervasive selection is also known to affect diversity at linked sites \citep{Sella2009} and could contribute to observed shifts away from equilibrium as well.

Historical and archaeological data are incompatible with substantial modern bottlenecks.  \emph{Megalonaias} from different regions show little differentiation across geographic regions \citep{Pfeiffer2018}.  Nevertheless, we still simulated a discrete 10 generation bottleneck representing a 5-fold population reduction (Figure \ref{ModBottleneck}).  Genetic diversity is largely unchanged, as finite samples taken from a larger population reflect equilibrium genetic diversity originally present. Such results contrast with the effects of ancient bottlenecks which alter Tajima's D as new mutations appear in populations over many generations and contribute to an excess of lower frequency alleles prior to reaching equilibrium.  Neither ancient nor modern bottlenecks could produce reductions in diversity similar to those observed in candidate regions for selective sweeps. While bottlenecks do increase the variance in $\pi$, we do not observe reductions below $\pi=0.0025$ in high recombination ($r=5\times10^{-8}$) or low recombination ($r=5\times10^{-9}$) regions. Thus, even implausible cases of strong bottlenecks that escaped historical observation cannot explain the sweep-like signals observed in natural populations.

\subsubsection*{Forward Simulations for Selection}
We then used SLiM v. 3.7 to perform forward simulations with tree sequence recording, then used recapitation to add neutral mutations onto simulated trees. We used simulation output to evaluate the impacts of very strong selection on $\pi$ and Tajima's $D$. We simulated a 10 Mb region for a population of 300,000 individuals setting recombination at $r=5\times10^{-8}$.  For simplicity and runtime we do not simulate full population growth scenarios implemented in faster coalescent simulations.  We first modeled population diversity for this scenario without selection to establish baseline diversity levels (Figure).  We then added a single new mutation favored by selection at 5 Mb. We used a fully dominant mutation ($h=1.0$) with very strong selection s=100, s=10, and s=0.1 under the SLiM fitness model of 1, $(1+hs)$, $(1+s)$ (Table \ref{SLimSims}).  Note that these selection coefficients correspond to s=0.99 and s=0.91 in relative fitness models using 1, $(1-hs)$ and $(1-s)$.  SLiM does not allow for simulation of strictly lethal selective regimes, but these selective coefficients of $s=10$ are consistent with up to 100-fold greater survival for individuals carrying an adaptive allele.  Given potentially lethal effects of pollution, pesticide use, host fish extinction, and other challenges facing Unionidae, extreme selective regimes are plausible and relevant for this species. 

\begin{figure}
\begin{center}
\includegraphics[scale=1]{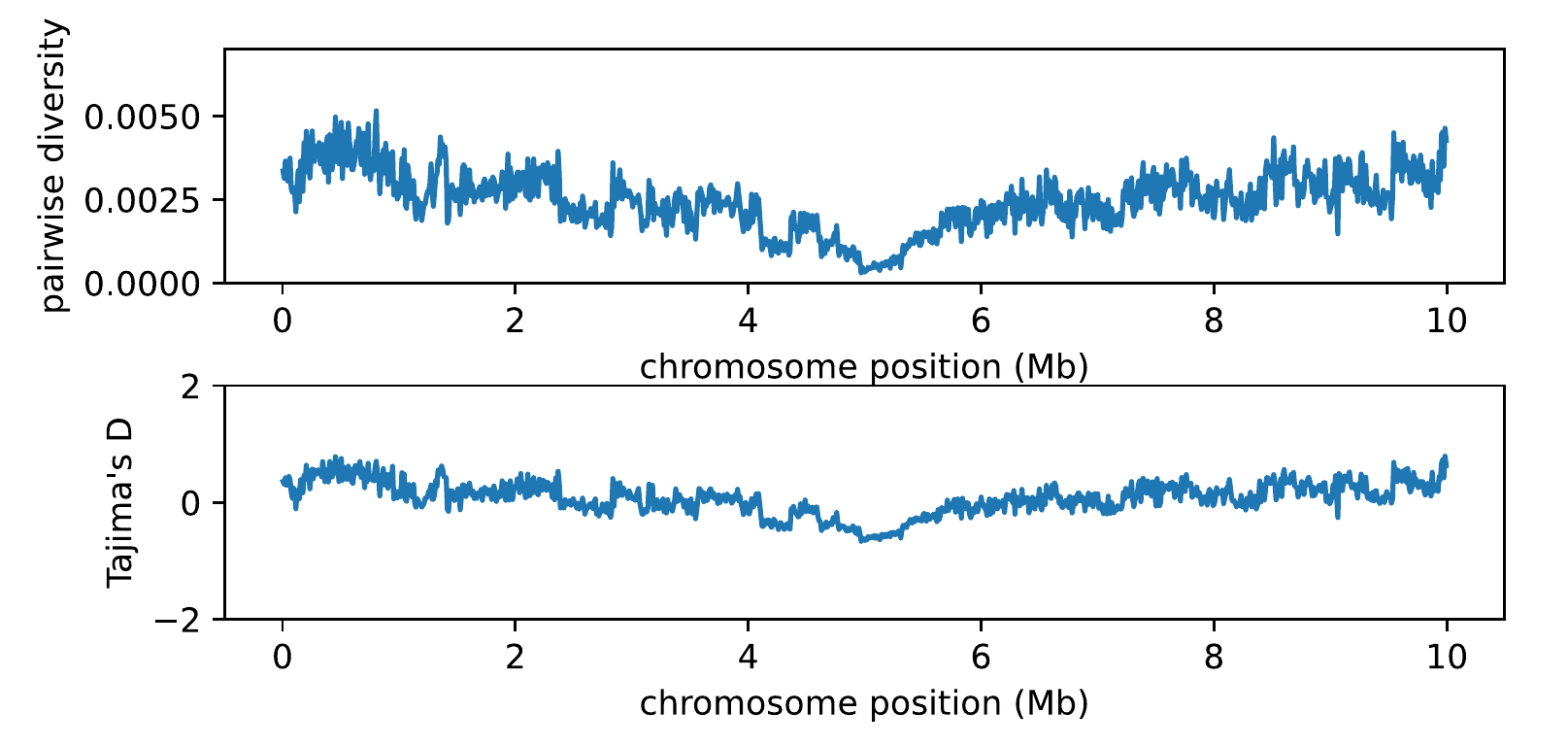}
\end{center}
\caption{Genetic diversity $\pi$ and Tajima's $D$ in response to a very strong selective sweep with $s=100$ after 20 generations. Simulations assume a population size of 300,000 individuals, with $r=5\times10^{-8}$.   \label{SweepR5-8s100g120}}
\end{figure}

\begin{figure}
\begin{center}
\includegraphics[scale=1]{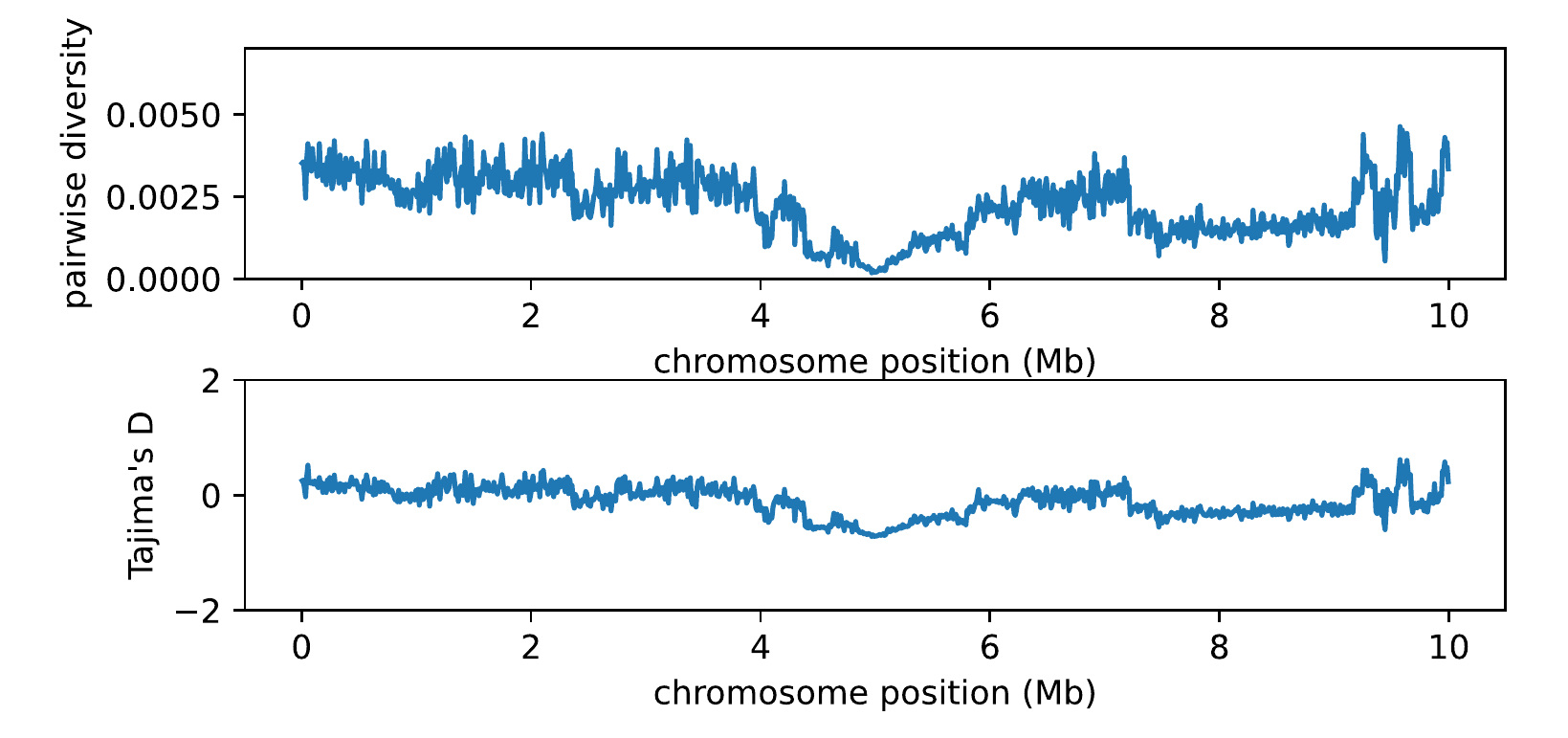}
\end{center}
\caption{Genetic diversity $\pi$ and Tajima's $D$ in response to a very strong selective sweep with $s=100$ after 50 generations. Simulations assume a population size of 300,000 individuals, with $r=5\times10^{-8}$. \label{SweepR5-8s100g150}}
\end{figure}

\begin{figure}
\begin{center}
\includegraphics[scale=1]{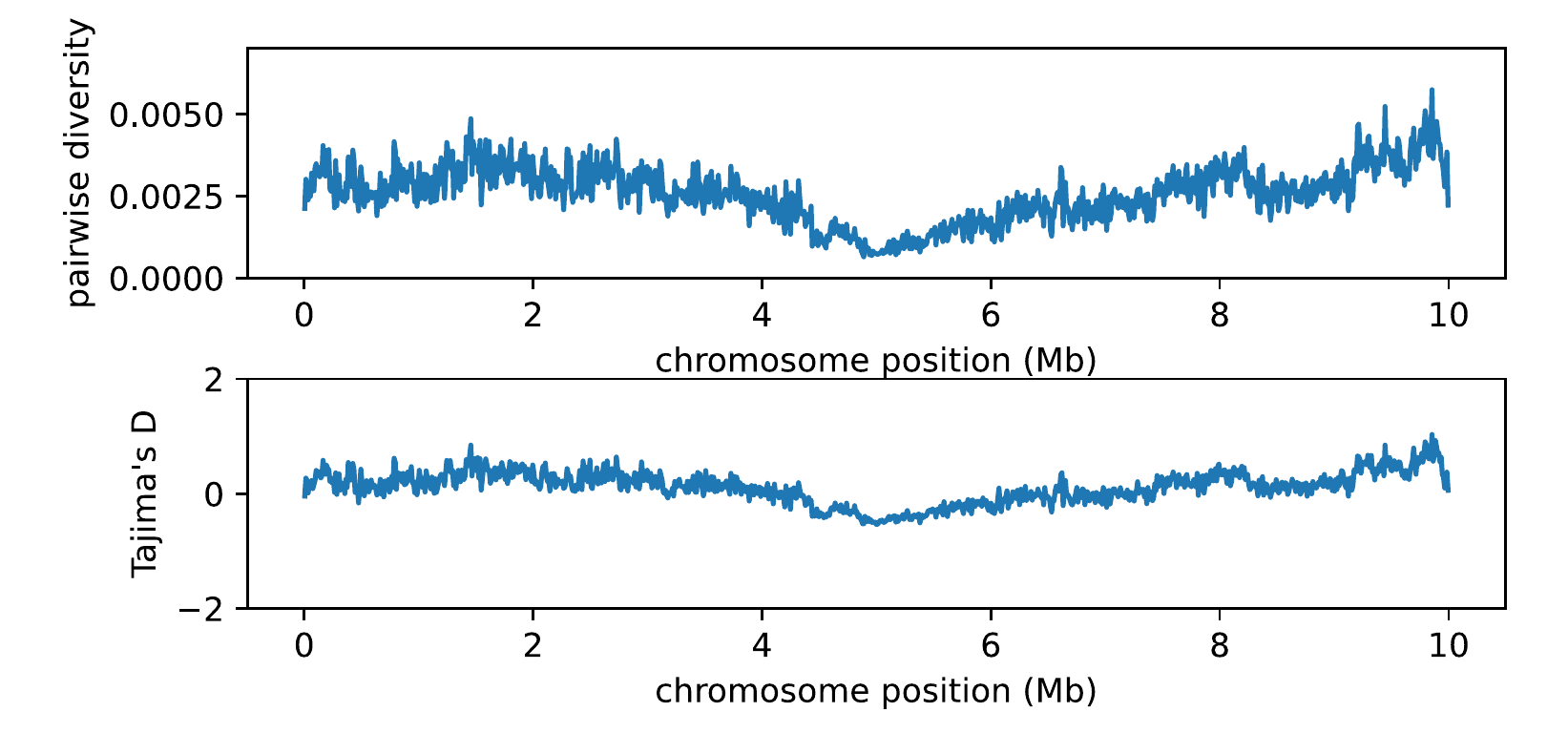}
\end{center}
\caption{Genetic diversity $\pi$ and Tajima's $D$ in response to a very strong selective sweep with $s=100$ after 15  generations. Simulations assume a population size of 300,000 individuals, with $r=5\times10^{-8}$.\label{SweepR5-8s100g115}}
\end{figure}

\begin{figure}
\begin{center}
\includegraphics[scale=1]{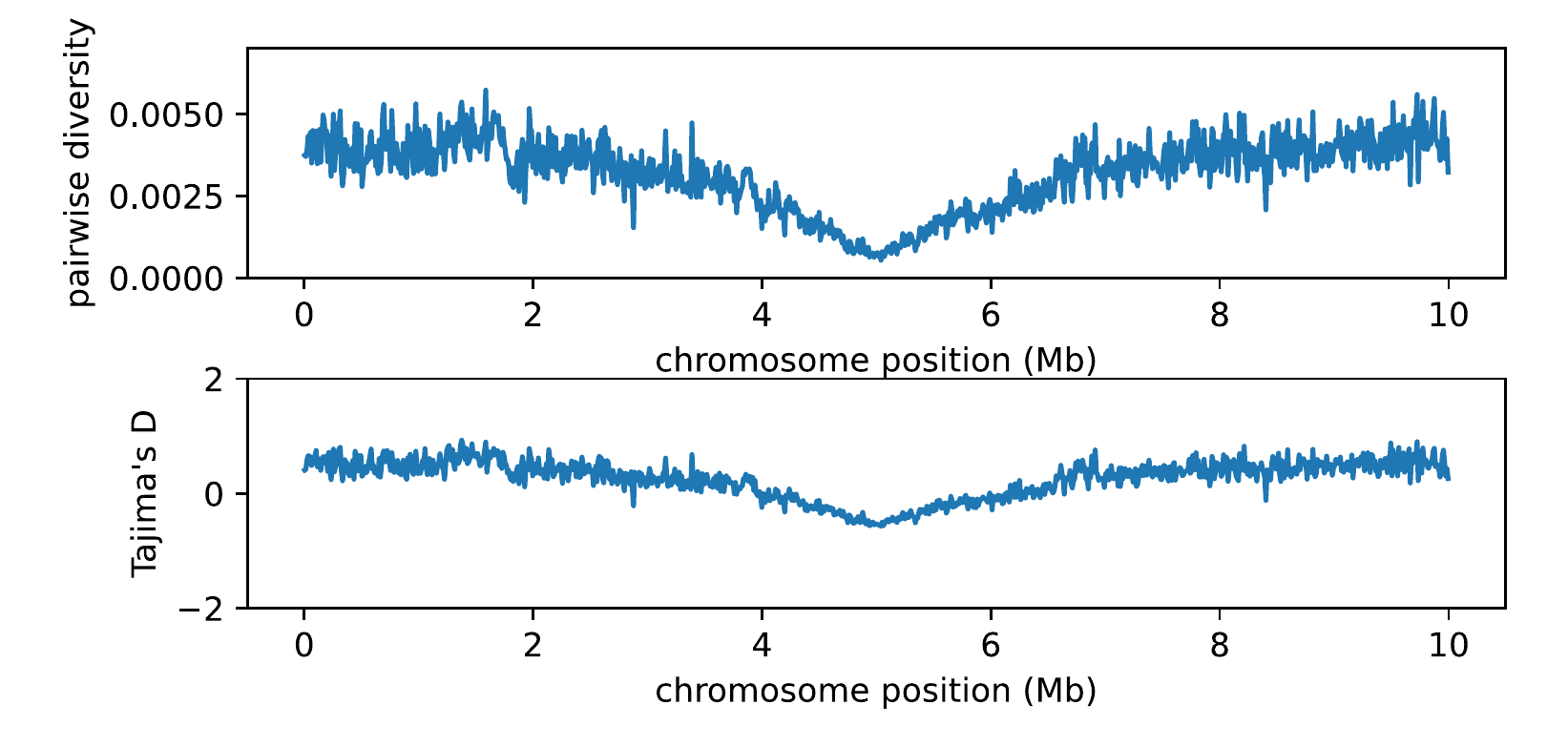}
\end{center}
\caption{Genetic diversity $\pi$ and Tajima's $D$ in response to a strong selective sweep with $s=10$ after 20 generations. Simulations assume a population size of 300,000 individuals, with $r=5\times10^{-8}$.  Diversity is reduced close to 0.0 in the 0.5 Mb region surrounding the selected locus. Diversity returns to near background levels towards the edges of the 10 Mb region.  \label{SweepR5-8s10g120}}
\end{figure}

\begin{figure}
\begin{center}
\includegraphics[scale=1]{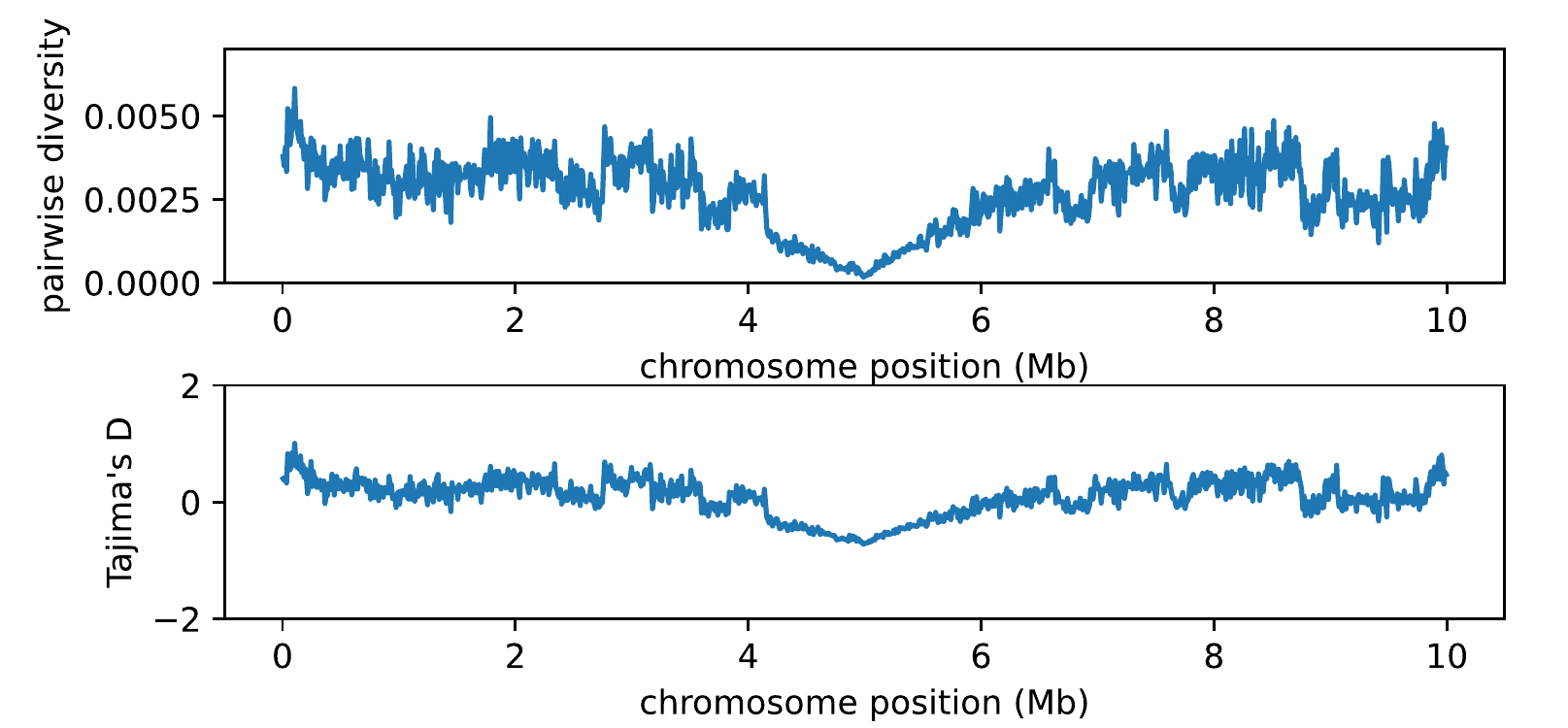}
\end{center}
\caption{Genetic diversity $\pi$ and Tajima's $D$ in response to a strong selective sweep with $s=10$ after 50 generations. Simulations assume a population size of 300,000 individuals, with $r=5\times10^{-8}$.  Diversity is reduced close to 0.0 for more than 1 Mb in the region surrounding the selected locus.  Diversity returns to near background levels towards the edges of the 10 Mb region. \label{SweepR5-8s10g150}}
\end{figure}

 With strong selection (s=10, s=100) genetic diversity drops 10-fold for 1-2 Mb in the central region around the selected locus in less than 20 generations, with more pronounced effects at 50 generations (Figure \ref{SweepR5-8s100g120}-\ref{SweepR5-8s10g150}).   Using simulated data as input for BCP, we can model the posterior mean and standard deviation for simulations.  These posterior distributions are used as models to establish goodness of fit tests compared with empirical data from scaffold 22 (a ~700kb contiguous scaffold from the largest sweep, Figure \ref{monomorphic}). Models with $s=100$ and $r=5\times10^{-8}$ are significantly better fit than $s=10$, $s=0.1$, or $r=5\times{10^-9}$ ($\chi^2 >40$ $df=1$ or $df=2$). We attempted to simulate selection at multiple sites, placing 3 selected sites 10kb apart.  In 1000 replicate simulations, none could produce sweeps on all 3 loci, due to interference.  Hence, multilocus selection is not a plausible explanation for the data.
 
\begin{figure}
\begin{center}
\includegraphics[scale=1]{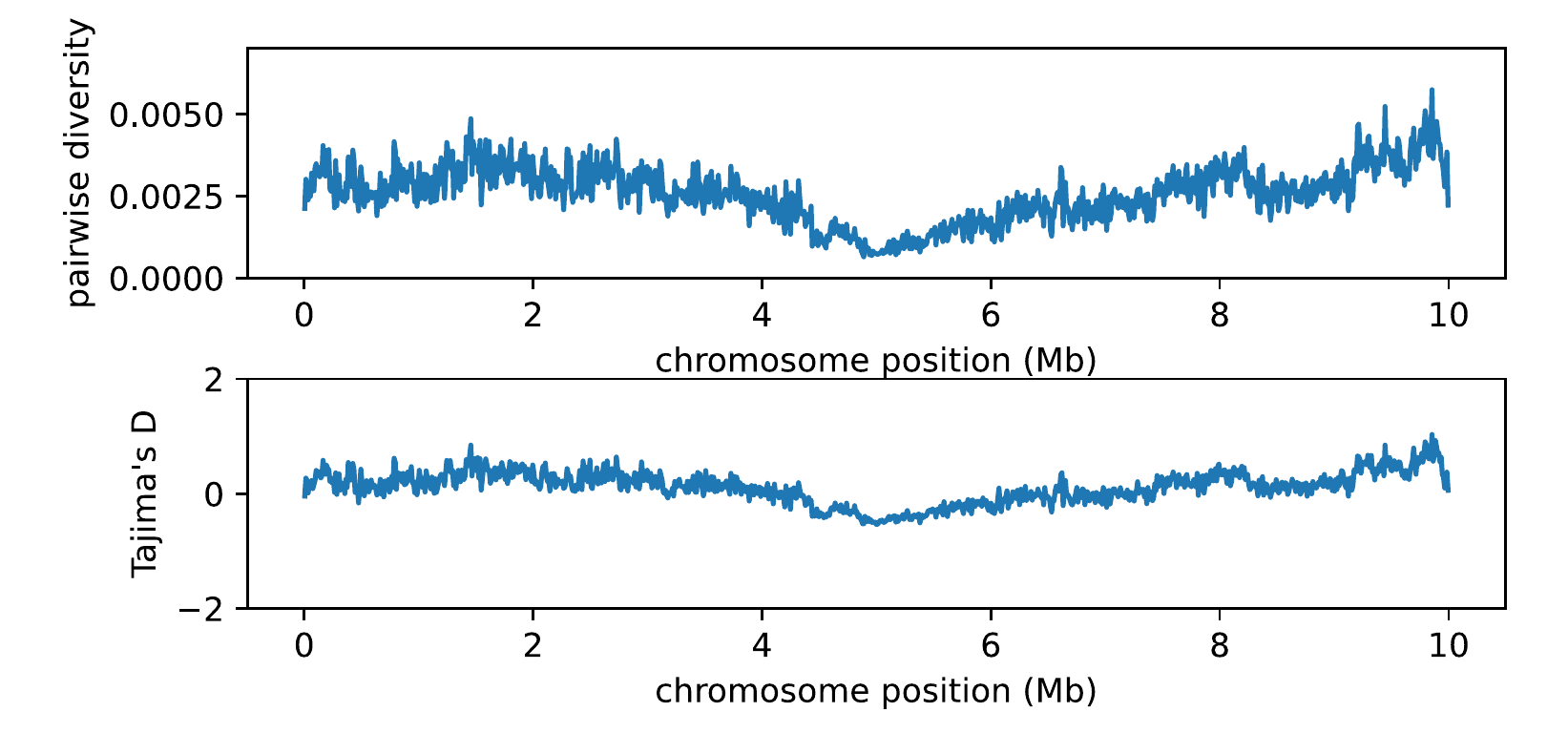}
\end{center}
\caption{Genetic diversity $\pi$ and Tajima's $D$ in response to a strong selective sweep with $s=10$ after 15  generations. Simulations assume a population size of 300,000 individuals, with $r=5\times10^{-8}$. \label{SweepR5-8s10g115}}
\end{figure}

\begin{figure}
\begin{center}
\includegraphics[scale=1]{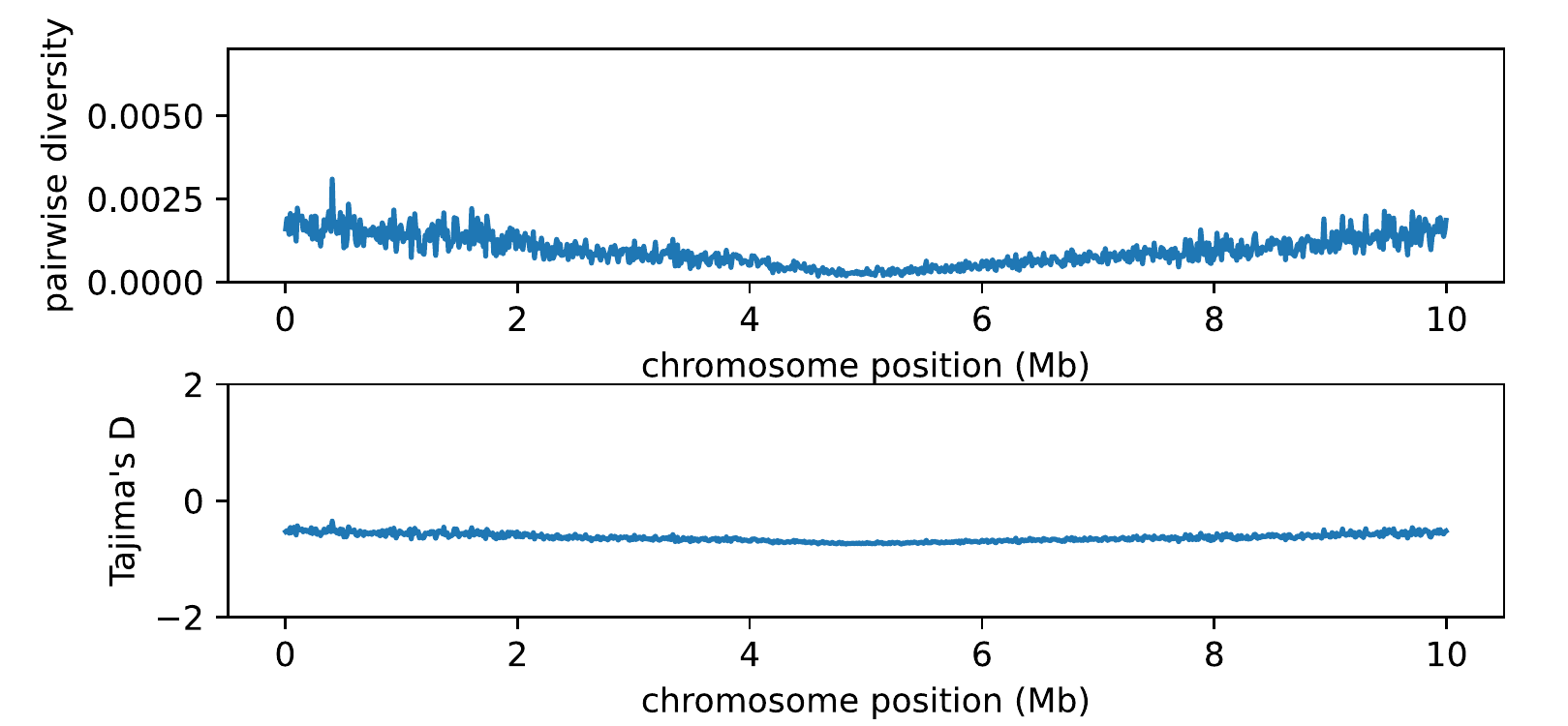}
\end{center}
\caption{Genetic diversity $\pi$ and Tajima's $D$ in response to a very strong selective sweep with $s=100$ after 50 generations. Simulations assume a population size of 300,000 individuals, with $r=5\times10^{-9}$. Under reduced recombination, background diversity is affected for the entire 10 Mb region and does not return to normal levels across the simulated contig, a pattern not observed in empirical data.  \label{SweepR5-9s100g150}}
\end{figure}

In simulations, genetic diversity adjacent to the sweep may be depressed to roughly half background levels in the 5 Mb region around the sweep. However, diversity returns to normal values near the edges of the 10 Mb region as recombination breaks linked haploytpes around the selected locus (Figure \ref{SweepR5-8s100g120}-\ref{SweepR5-8s10g150}).  At 15 generations, the reduction in diversity is observable but more modest (Figure \ref{SweepR5-8s100g115}, Figure \ref{SweepR5-8s10g115}).  Selection on standing variation at frequencies above $p=0.01$ would establish deterministic sweeps even more quickly \citep{Hermisson2005}, but is not easy to simulate (as discussed thoroughly in the SLiM Manual).  Under recombination reduction to $r=1\times10^{-8}$ or $r=5\times10^{-9}$, diversity is still reduced around the selected locus, but background genetic diversity is severely depressed across the entire 10 Mb region, inconsistent with empirical data (e.g. Figure \ref{SweepR5-9s100g150}).  Hence, we do not believe that pervasive recombination suppression is compatible with data available for \Mnerv.
 
\begin{figure}
\begin{center}
\includegraphics[scale=1]{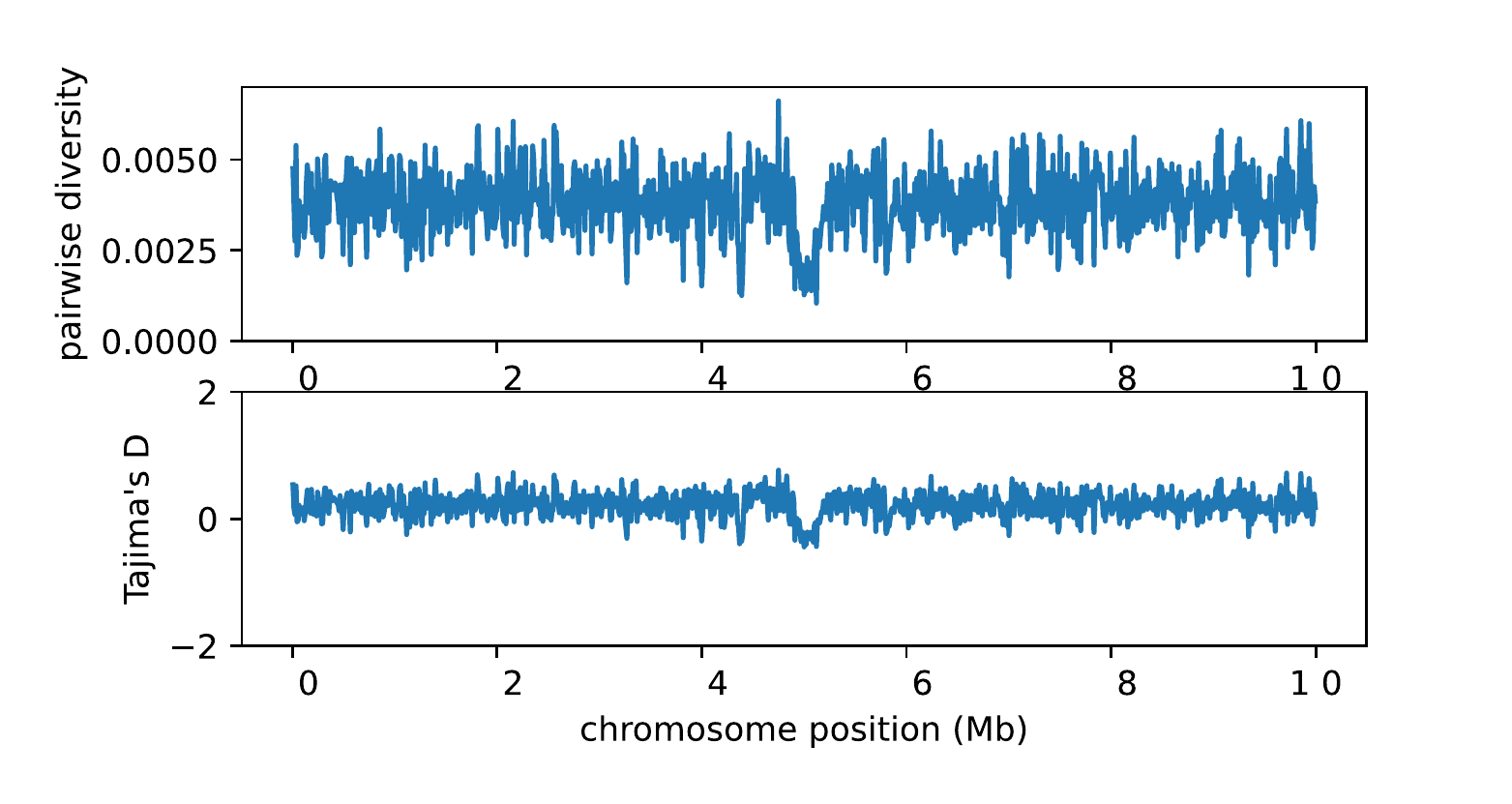}
\end{center}
\caption{Genetic diversity $\pi$ and Tajima's $D$ in response to a very strong selective sweep with $s=0.1$ after 300 generations under normal recombination. Simulations assume a population size of 300,000 individuals, with $r=5\times10^{-8}$. These simulations show the classic v-like pattern for selective sweeps rather than extended blocks with near zero diversity.  \label{SweepR5-8s0pt1g300}}
\end{figure}

\begin{figure}
\begin{center}
\includegraphics[scale=1]{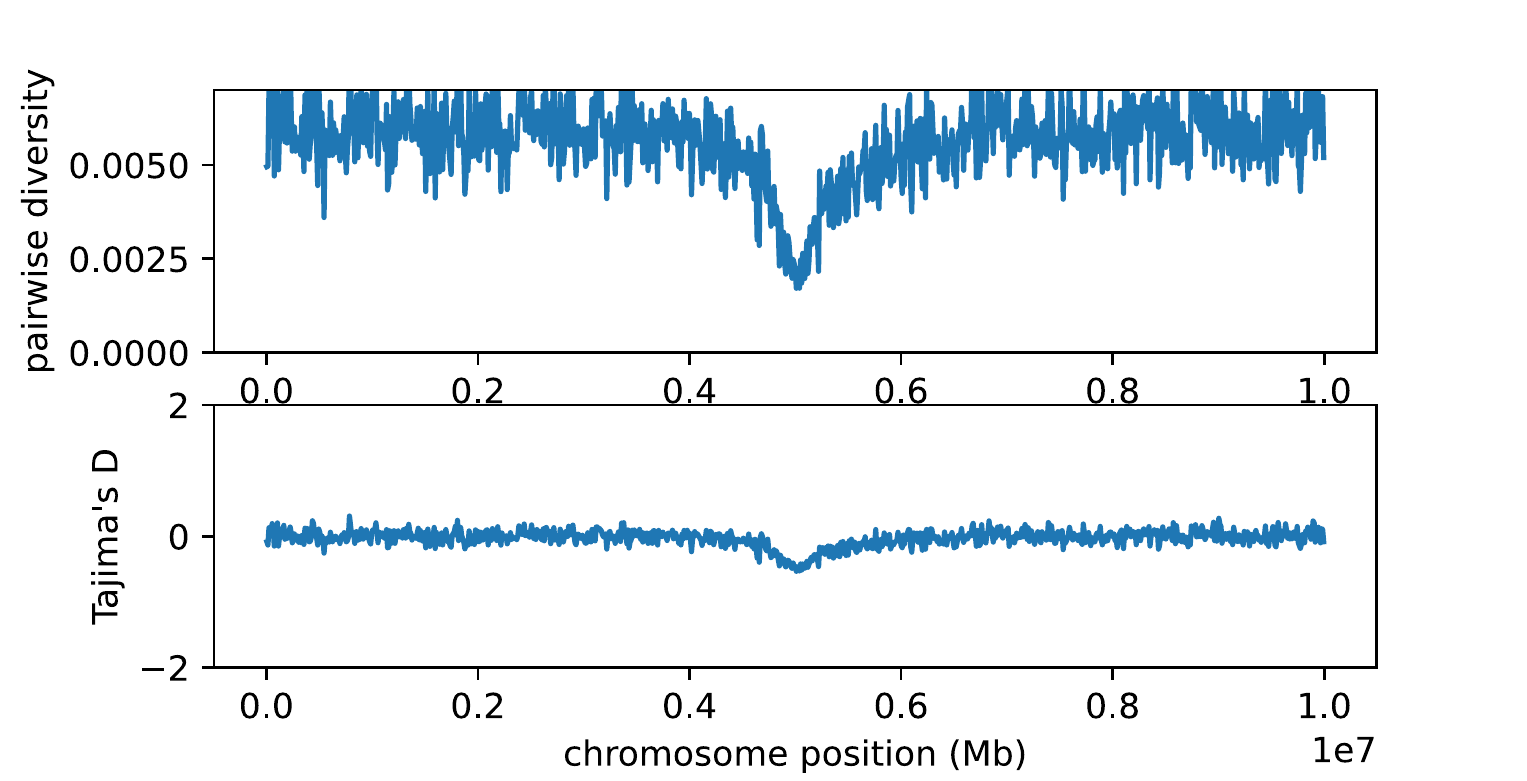}
\end{center}
\caption{Genetic diversity $\pi$ and Tajima's $D$ in response to a weak selective sweep with $s=0.1$ after 300 generations under moderately reduced recombination. Simulations assume a population size of 300,000 individuals, with $r=1\times10^{-8}$.  These simulations show the classic v-like pattern for selective sweeps rather than extended blocks with near zero diversity but with a wider scope than in normal recombination.  \label{SweepR1-8s0pt1g300}}
\end{figure}

\begin{figure}
\begin{center}
\includegraphics[scale=1]{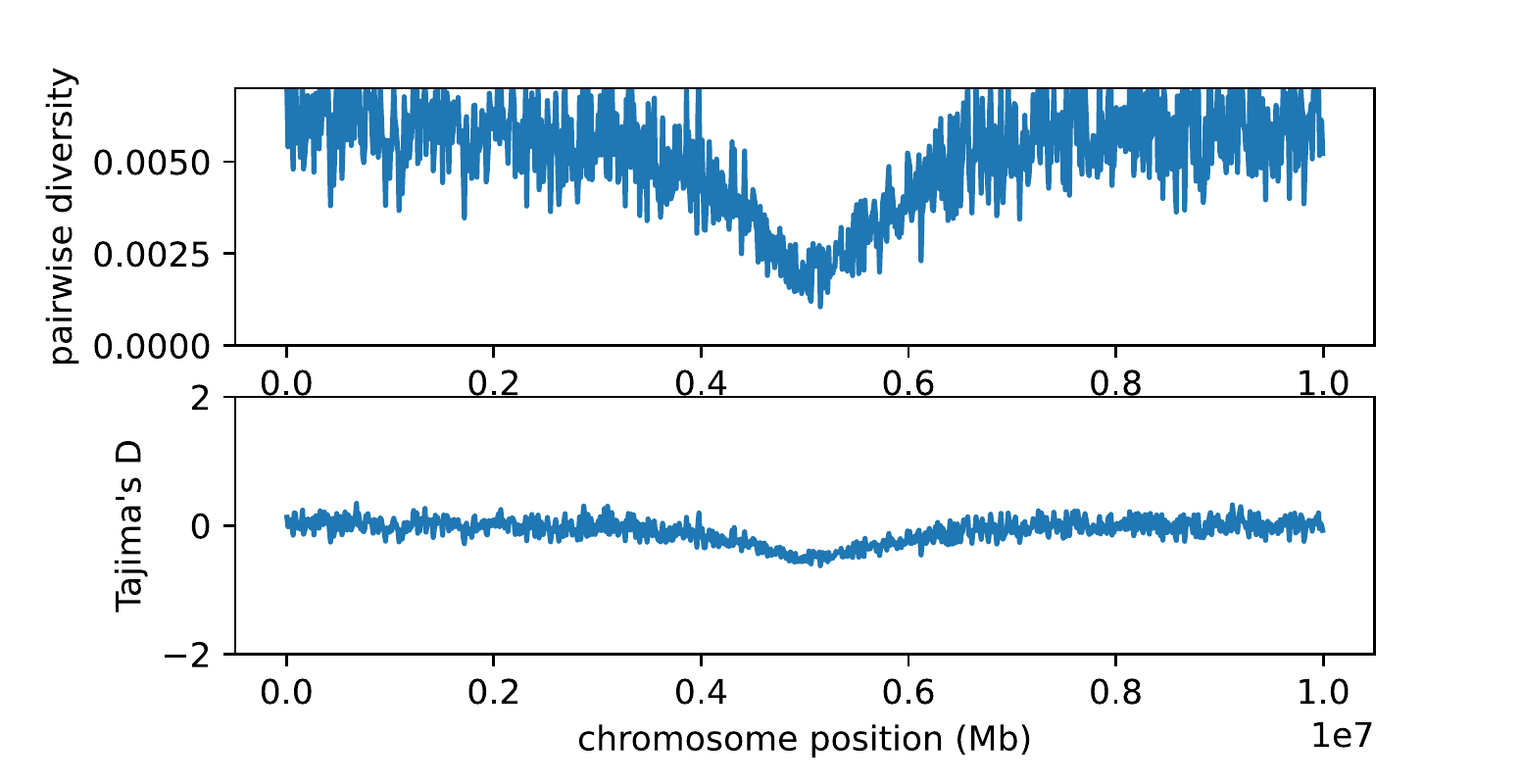}
\end{center}
\caption{Genetic diversity $\pi$ and Tajima's $D$ in response to a very strong selective sweep with $s=100$ after 50 generations under moderately reduced recombination. Simulations assume a population size of 300,000 individuals, with $r=1\times10^{-8}$.  These simulations show the classic v-like pattern for selective sweeps rather than extended blocks with near zero diversity, though the pattern is broader than in normal recombination.   \label{SweepR5-9s0pt1g300}}
\end{figure}

Under lower selection coefficients ($s=0.1$) and normal recombination ($r=5\times10^{-8}$), the selected allele was lost due to stochastic fluctuations affecting a singleton allele in the first 4 out of 5 initial simulation attempts, consistent with theory \citep{Hermisson2005}.  When a deterministic sweep did establish, these simulations began to show sweep-like signals around 300 generations, but not by 200 generations.  Genetic diversity approaches zero near the surrounding locus, but produced a classic v-shaped pattern as recombination breaks apart linkage during the sweep (Figure \ref{SweepR5-8s0pt1g300}-\ref{SweepR5-9s0pt1g300}).   With reduced recombination ($r=5\times10^{-9}$) 19 out of 20 simulations resulted in loss of the allele.  These simulations with mild selection do not produce the extensive megabase sized tracts with reduced in genetic diversity that we observe in empirical data.   Hence, selective regimes with $s=0.1$ are not sufficient to produce the extreme population genomic signals we have observed, even over long timescales. 

In light of these simulations, we suggest that scenarios of very strong selection are sufficient and necessary to explain reductions in genetic diversity observed in the natural population of \Mnerv.

\clearpage
\begin{table}
\begin{center}
\footnotesize
\caption{ \label{Specimens} Specimens of \Mnerv {} from Pickwick Lake Oct 2018 }
\begin{tabular}{lrrrrrrr}
Number & Length (cm) & Depth (cm) & Height (cm)  & Ring Count & Gravid & Sequencing Status\\
\hline
1 & 12.3 &5.2  & 8.4 & 13 & Yes&Sequenced \\ 
2 & 10.0 & 4.4  & 7.4  & 11 & Yes & Sequenced \\ 
3 & 15.8  & 5.7    & 11.0  & 11 & Yes & Reference Genome\\ 
4 &  10.9 & 4.7 &8.2 & 10 & Yes &Sequenced \\ 
5 & 10.6 & 4.6 & 7.9 & 8 & Yes &Sequenced \\ 
6 & 11.4 & 4.8  & 8.4  & 9 & Yes &Sequenced \\
7 & 9.9 & 4.1  & 7.3 & 7 & Yes & Sequenced \\
8 & 12.4 &  4.6 &9.0 & 10 & No  & Sequenced \\
9 & 10.4 & 3.8 & 7.1 & 6&  Yes  & Sequenced\\
10 & 10.4 & 4.4 & 7.5 & 11  & No &Sequenced \\
11 & 12.1 & 4.9 & 8.5 & 12 & Yes& Sequenced \\
12 & 10.7  & 4.2  & 7.8  & 10 &  Yes &Sequenced \\
13 &  10.4 & 4.2  & 8.1  & 8  & Yes& Sequenced \\
14 & 10.9 & 4.7 & 8.6 &  8 & Yes &Not sequenced \\
15 & 11.9 & 5.8 & 11.9 & 11 & Yes & Not sequenced \\
16 & 8.5  & 2.5 &  5.0 & 8 & No  & Sequenced \\
\hline
\end{tabular}
\end{center}
\end{table}

\clearpage
\begin{table}
\begin{center}
\caption{Diversity Metrics}
\begin{tabular}{llr}
\hline
$\pi$ & mean & 0.005447792 \\
& median & 0.005031502 \\
& st. dev. & 0.002679202 \\

Tajima's $D$ & mean & -1.160924 \\
& median & -1.220574 \\
& st. dev. & 0.8012326 \\
\hline
\end{tabular}
\label{Pgen}
\end{center}
\end{table}
%
%

\clearpage
\begin{table}
\begin{center}
\caption{Demographic simulations}
\footnotesize
\begin{tabular}{llllll}

Scenario & growth start & growth rate & duration   & Ratio \\
\hline
Ancient Expansion  & 2000 & 0.25 & 10 & 10X\\
 & 5000* &  0.25 & 10  & 10X\\
  & 10000 & 0.25 & 10  & 10X \\
Continuous Expansion & 2000 & 0.00125 &  - & 12X  \\
& 5000* & 0.005 & - & 12X \\
& 10000 & 0.00025 & - & 12X \\
Modern Bottleneck & 10 & instant & 10 & 5X \\
\hline
*Best fit models for for whole genome background diversity.
\end{tabular}
\label{PopExp}
\end{center}
\end{table}

\clearpage
\begin{table}
\begin{center}
\caption{Simulation parameters}
\begin{tabular}{llr}

Recombination $r$ & Selection $s$  \\
\hline
$5\times10^{-8}$ & 0.1 \\
 & 10 \\
 & 100* \\
$1\times10^{-8}$ & 0.1 \\
 & 10 \\
 & 100 \\
$5\times10^{-9}$ & 0.1 \\
 & 10 \\
 & 100 \\
\hline
*Best fit model for for scaffold 22 test data.
\end{tabular}
\label{SLimSims}
\end{center}
\end{table}

\clearpage
\begin{table}
\begin{center}
\caption{ \label{GenesInSweeps} Genes in Selective Sweeps }
\tiny
\begin{tabular}{lll}
\hline
Apoptosis inhibitor & 18 & g7640, g37204, g40603, g9485, g65154, g13778, g83829,  g15106, g2052, g2174, g103700, g19660, \\
& &g2931, g23549, g130064, g26500; g77619 (65 kDa Yes-associated protein); g28882 (APAF1-interacting protein) \\
Stress Tolerance & 17 &g75456, g75457, g81615, g81619, g97141, g110382, g110383 (Hsp70); \\
& &g49053, g72756, g85437, g100073 (Early growth response protein); g15855, g96078,g96079 (Carnitine metabolism); \\
& & g115821, g115825, g115827 (Ectonucleoside triphosphate diphosphohydrolase 7) \\
Shell Formation & 10 & g41573 (Chitin synthase 3); g32, g36, g48649 (sodium bicarbonate exchanger); g2794 (Perlucin);\\
& & g37568 (Anoctamin); g56763, g72205, g87285, g3813 (Calmodulin)  \\
Development & 9 & g8580 (Bone morphogenetic protein 3);g87703, g73099, g75719, g104521, g120171, g30182 (Homeobox); \\
& &g65282, g65283 (Wnt) \\
Detox & 8 & g62671, g62674  (Cytochrome b5 reductase 4); g59044 (Rieske domain-containing protein); \\
& &g81499 (Cytochrome P450); g13536 (Dimethylaniline monooxygenase); \\
& &g43545 (Flavin-containing monooxygenase); g77388 (Peroxiredoxin); \\
& &g73090 (Protoporphyrinogen oxidase - target of herbicides) \\
Mitochondria & 8 &  g22770 (Atlastin); g13794, g13795, (ADP,ATP carrier protein 1,); g84734(ATP synthase subunit delta);\\
& & g15674 (Fumarylacetoacetate hydrolase); g74352, g74355 (Isocitrate dehydrogenase); \\
& &g105152 (Mitochondrial fission factor), g39152 (Nuclease EXOG, mitochondrial)\\
Parasite-Host & 6 & g87422 (Plasminogen); g84442 (Pleckstrin); g100841 (Fibronectin); \\
& &g83864 (Multimerin-1); g57648, g104196 (von Willebrand factor)\ \\
DNA repair &6  & g2915, g2920, g2909 (DNA mismatch repair protein); \\
& &g35257, g35258, g74360 (Werner syndrome helicase) \\
\hline
\end{tabular}
\end{center}
\end{table}
\clearpage

\clearpage
\begin{table}
\begin{center}
\caption{ \label{TEType} Polymorphic TE Insertions in \Mnerv }
\begin{tabular}{rl}
\hline
\emph{Polinton} &1701 \\
\emph{Gypsy} &1570 \\
\emph{RTE} &171 \\
\emph{L2} &138 \\
\emph{Nimb} &135 \\
\emph{Penelope} &130 \\
\emph{Daphne} &80 \\
\emph{Neptune} &78 \\
\emph{CR1} &71 \\
\emph{SAT} &69 \\
All Other & 828 \\
\hline
 Total & 4971 \\
\hline
\end{tabular}
\end{center}
\end{table}

\clearpage
\begin{figure}
\begin{center}
\includegraphics[scale=1]{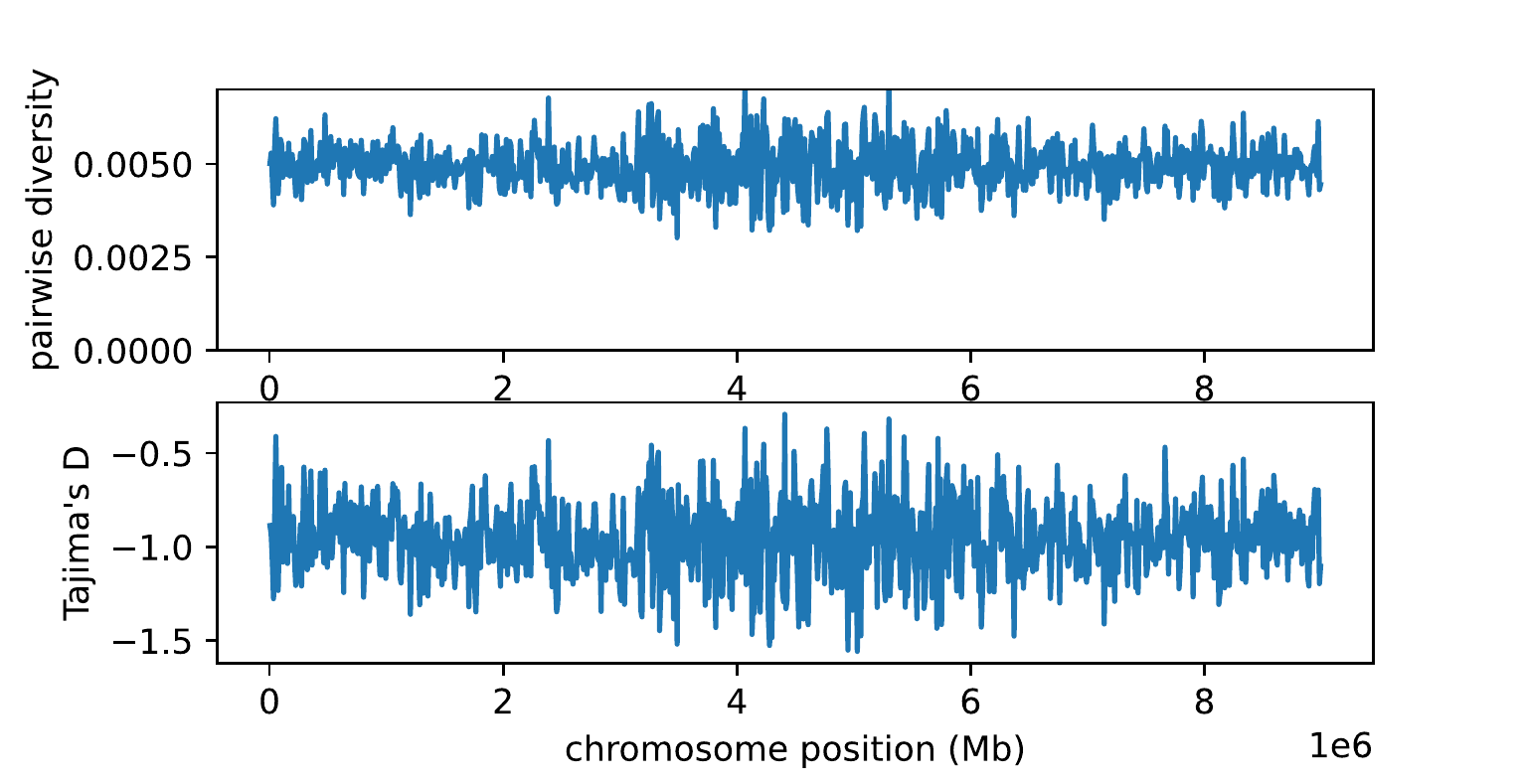}
\end{center}
\caption{Genetic diversity $\pi$ and Tajima's $D$ under simulated 12-fold population growth at a rate of 1.0005 per generation over 5000 generations. Simulations show highly recombining regions ($r=5\times10^{-8})$ from 0-3 Mb and 6-9 Mb, and lowly recombining regions ($r=5\times10^{-9}$) from 3-6 Mb. \label{PopGrowth5000}}
\end{figure}

\clearpage
\begin{figure}
\begin{center}
\includegraphics[scale=1]{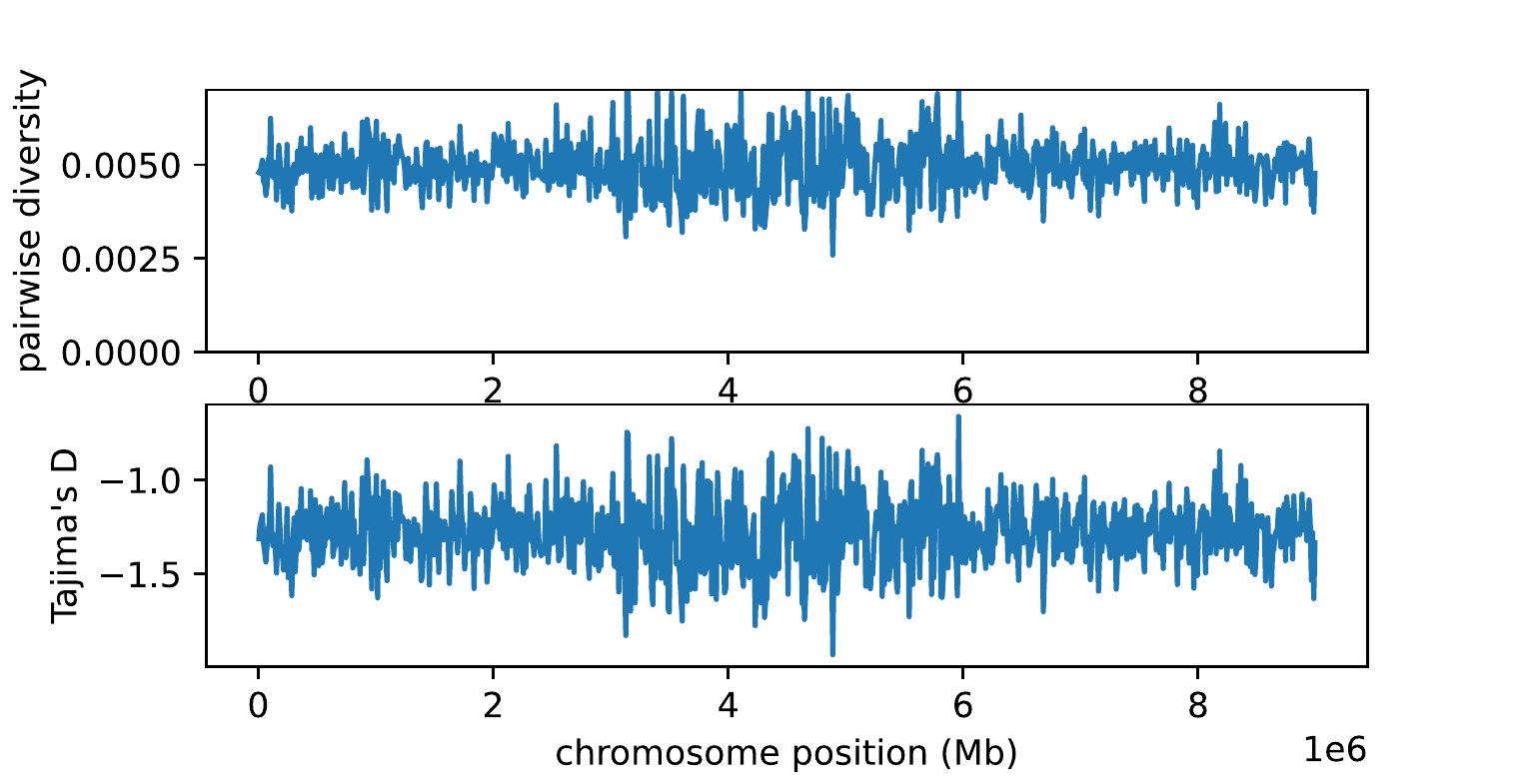}
\end{center}
\caption{Genetic diversity $\pi$ and Tajima's $D$ under simulated 12-fold population growth at a rate of 1.00025 per generation over 10000 generations. Simulations show highly recombining regions ($r=5\times10^{-8}$) from 0-3 Mb and 6-9 Mb, and lowly recombining regions ($r=5\times10^{-9}$) from 3-6 Mb.  \label{PopGrowth10000}}
\end{figure}

\clearpage
\begin{figure}
\begin{center}
\includegraphics[scale=1]{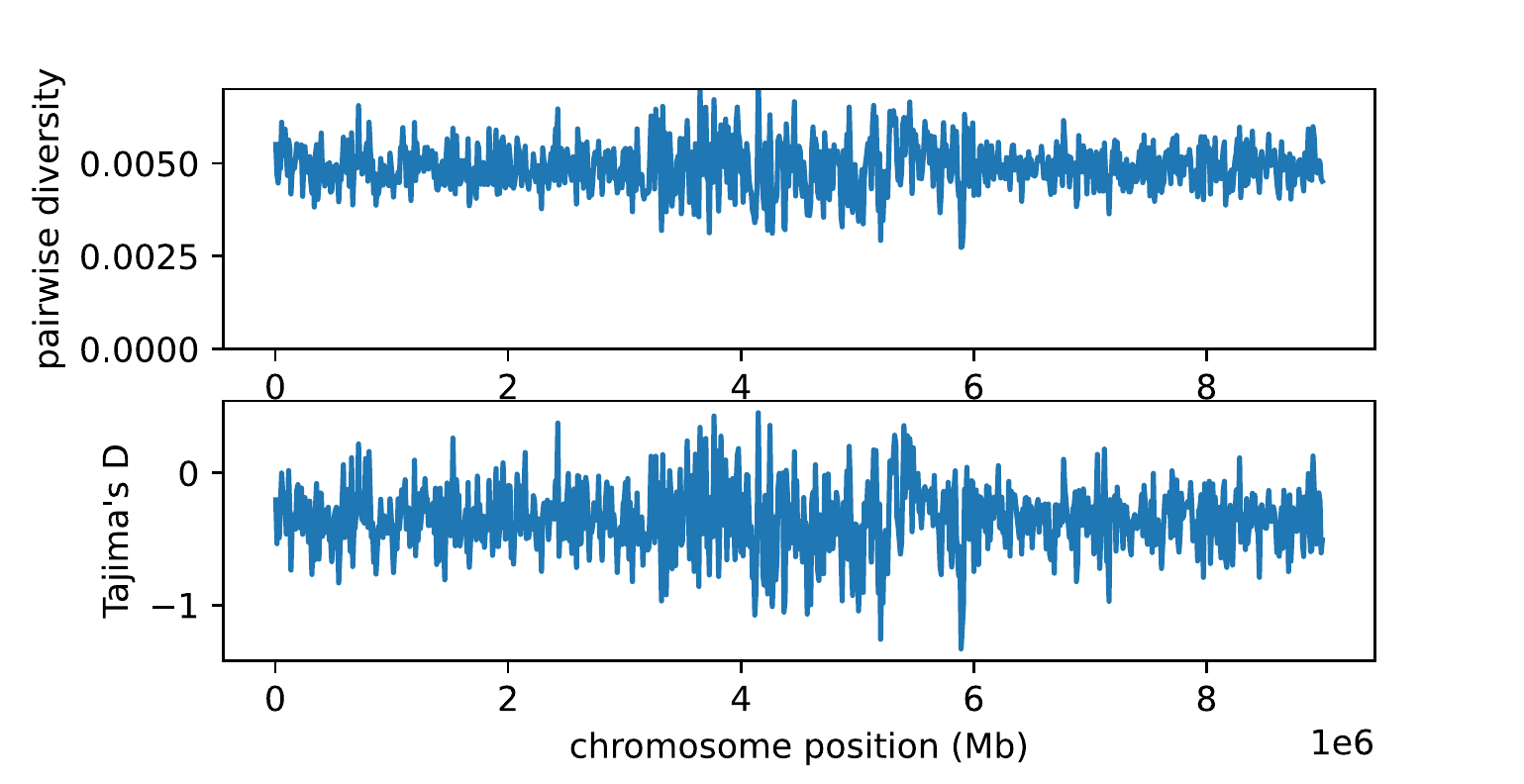}
\end{center}
\caption{Genetic diversity $\pi$ and Tajima's $D$ under simulated 12-fold population growth at a rate of 1.0025 per generation over 1000 generations. Simulations show highly recombining regions ($r=5\times10^{-8}$) from 0-3 Mb and 6-9 Mb, and lowly recombining regions ($r=5\times10^{-9}$) from 3-6 Mb.  \label{PopGrowth1000}}
\end{figure}

\clearpage
\begin{figure}
\begin{center}
\includegraphics[scale=1]{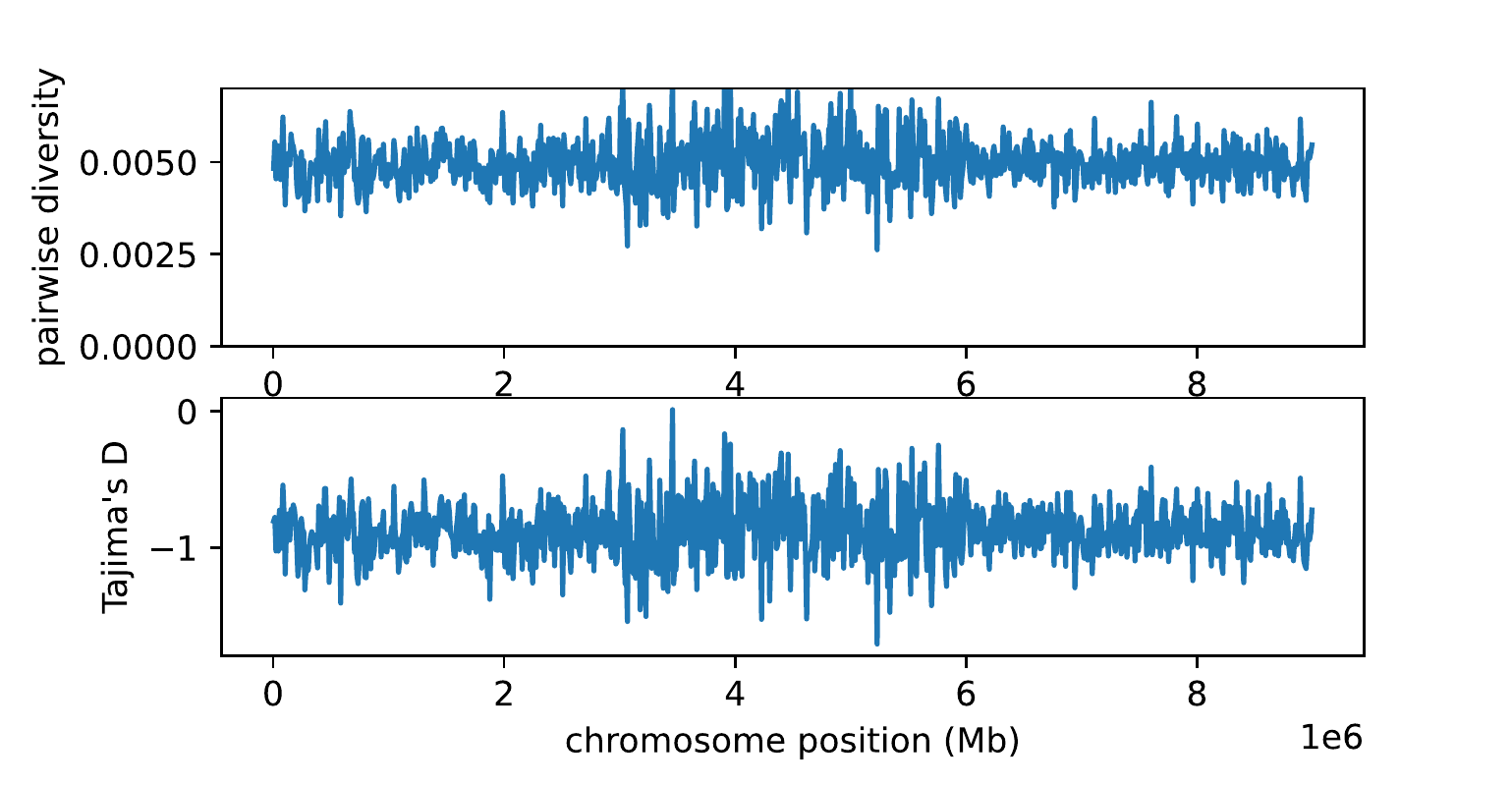}
\end{center}
\caption{Genetic diversity $\pi$ and Tajima's $D$ under simulated 12-fold population growth at a rate of 1.0005 per generation over 1000 generations with a 5-fold population bottleneck 10 generations in the past followed by population recovery. Simulations show highly recombining regions ($r=5\times10^{-8}$) from 0-3 Mb and 6-9 Mb, and lowly recombining regions ($r=5\times10^{-9}$) from 3-6 Mb.  \label{PopGrowth1000}}
\end{figure}

\clearpage

\clearpage

\clearpage

\clearpage

\clearpage

\clearpage

\clearpage

\clearpage
\begin{figure}
\begin{center}
\includegraphics[scale=1]{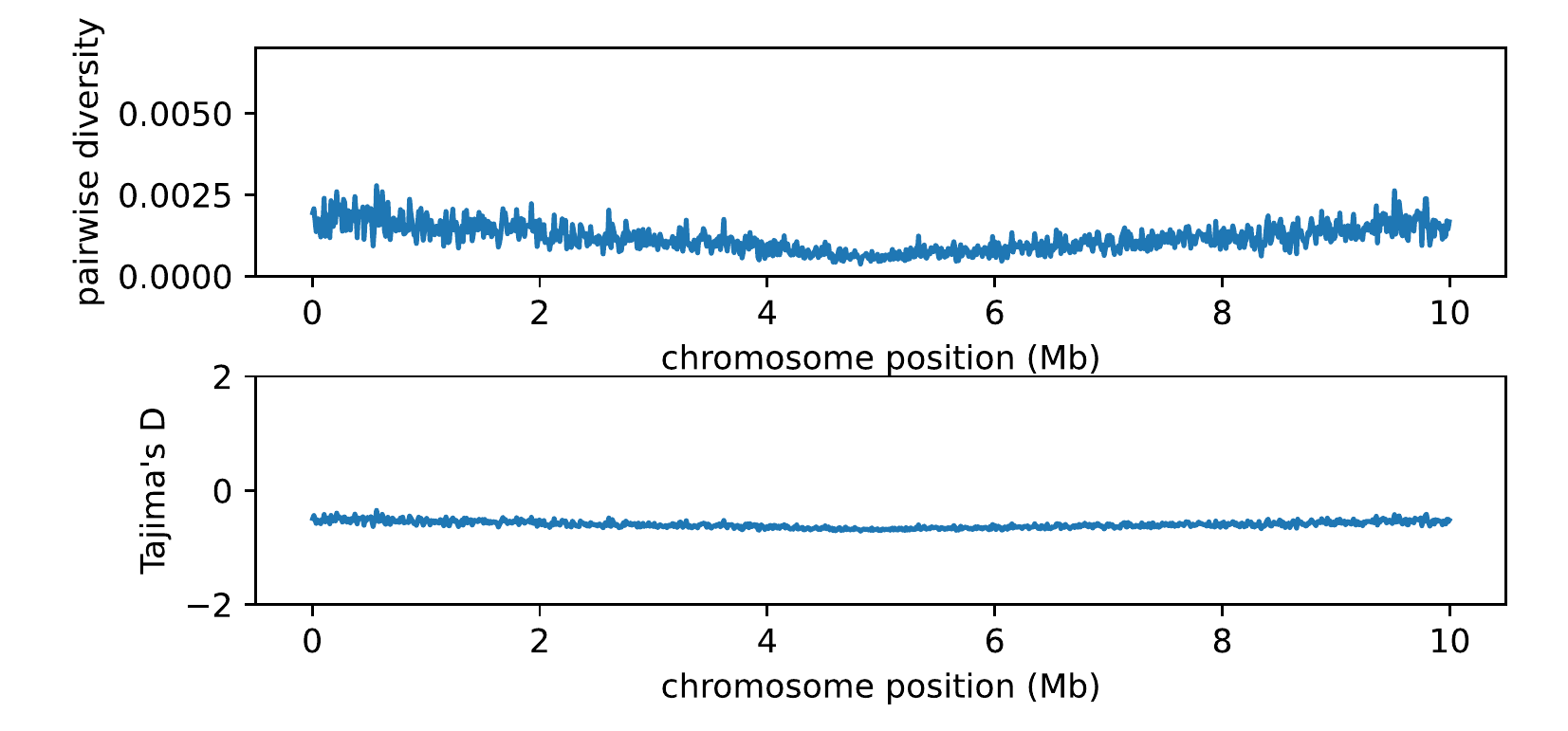}
\end{center}
\caption{Genetic diversity $\pi$ and Tajima's $D$ in response to a very strong selective sweep with $s=100$ after 20 generations in a genome with low recombination. Simulations assume a population size of 300,000 individuals, with $r=5\times10^{-9}$. Under reduced recombination, background diversity is affected for the entire 10 Mb region and does not return to normal levels across the simulated contig, a pattern not observed in empirical data.   \label{SweepR5-9s100g120}}
\end{figure}

\clearpage

\clearpage
\begin{figure}
\begin{center}
\includegraphics[scale=1]{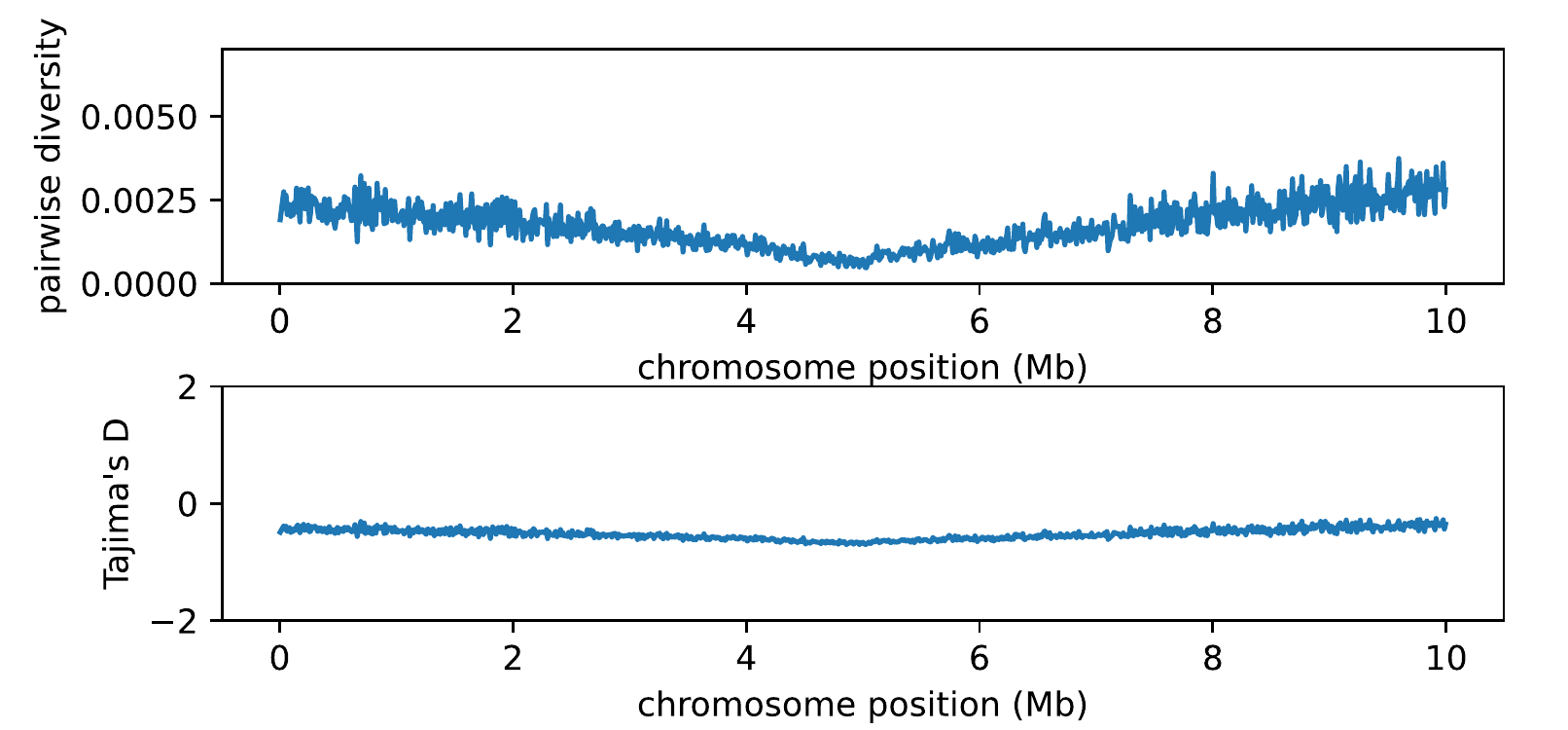}
\end{center}
\caption{Genetic diversity $\pi$ and Tajima's $D$ in response to a very strong selective sweep with $s=100$ after 20 generations under moderately reduced recombination. Simulations assume a population size of 300,000 individuals, with $r=1\times10^{-8}$. Under reduced recombination, background diversity is affected for the entire 10 Mb region and does not return to normal levels across the simulated contig, a pattern not observed in empirical data.   \label{SweepR1-8s100g120}}
\end{figure}

\clearpage
\begin{figure}
\begin{center}
\includegraphics[scale=1]{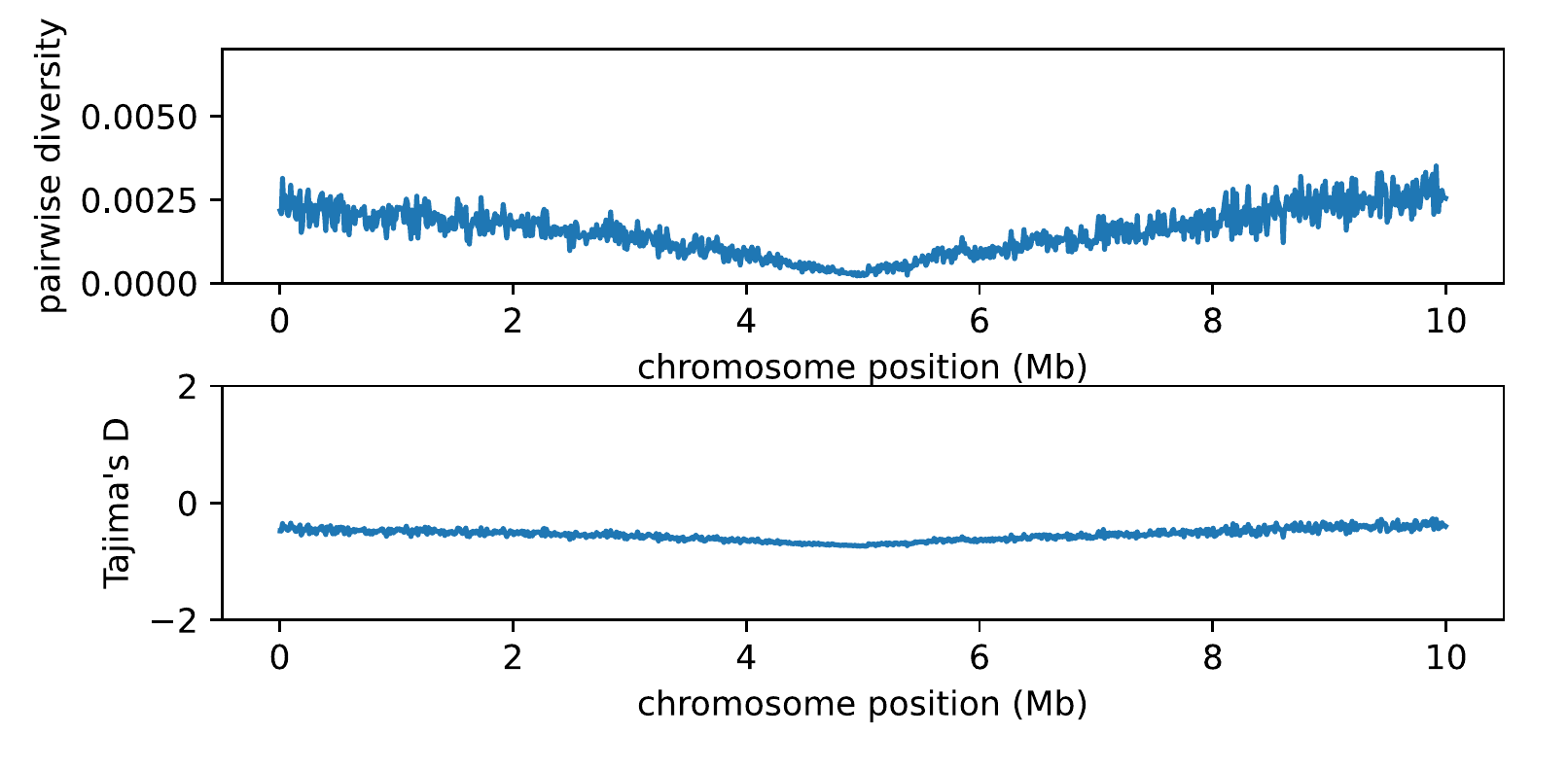}
\end{center}
\caption{Genetic diversity $\pi$ and Tajima's $D$ in response to a very strong selective sweep with $s=100$ after 50 generations under moderately reduced recombination. Simulations assume a population size of 300,000 individuals, with $r=1\times10^{-8}$. Under reduced recombination, background diversity is affected for the entire 10 Mb region and does not return to normal levels across the simulated contig, a pattern not observed in empirical data.  \label{SweepR1-8s100g150}}
\end{figure}
\clearpage

\clearpage
\begin{figure}
\begin{center}
\includegraphics[scale=1]{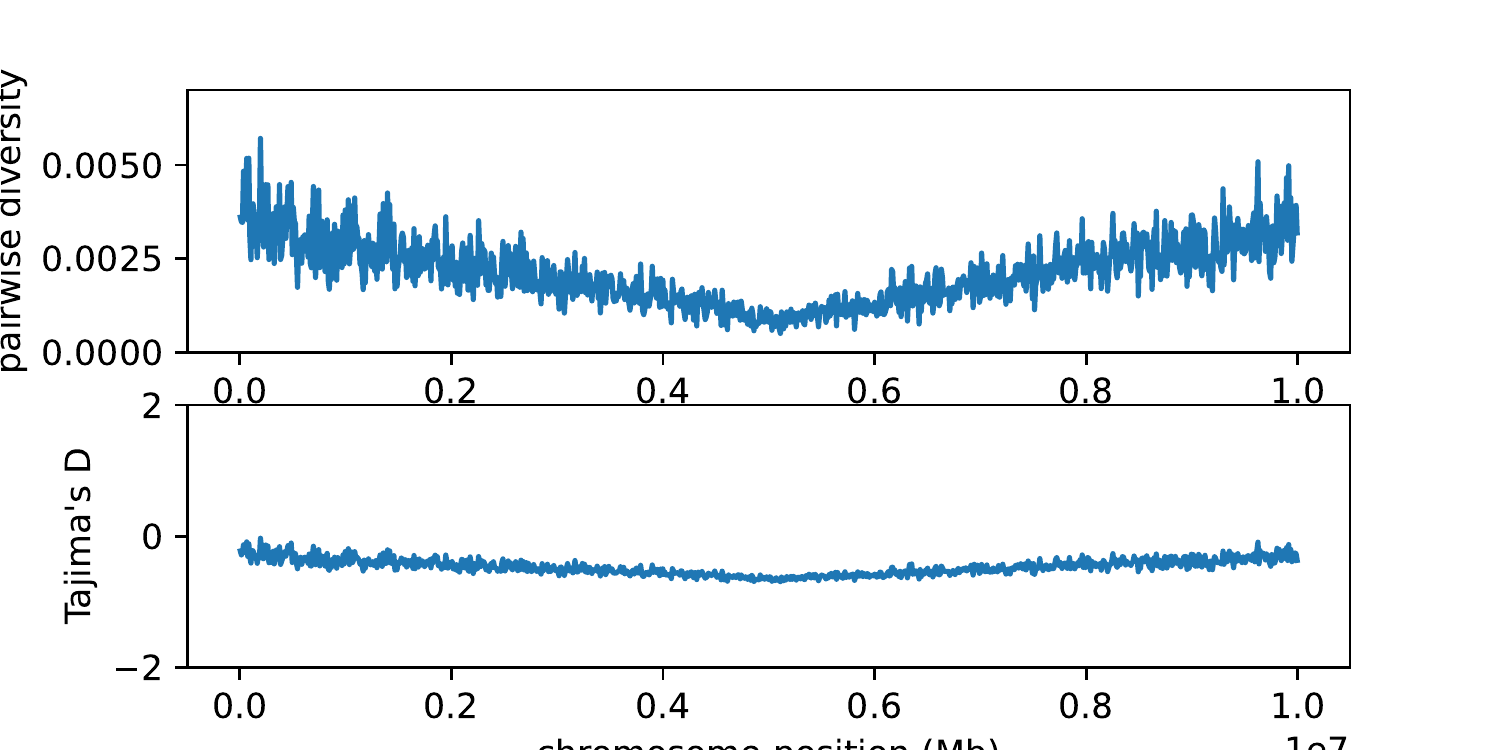}
\end{center}
\caption{Genetic diversity $\pi$ and Tajima's $D$ in response to a very strong selective sweep with $s=10$ after 20 generations under moderately reduced recombination. Simulations assume a population size of 300,000 individuals, with $r=1\times10^{-8}$. Under reduced recombination, the locus of the sweep does not move as close to zero diversity and background diversity is affected for the entire 10 Mb region and does not return to normal levels across the simulated contig, a pattern not observed in empirical data.   \label{SweepR1-8s100g120}}
\end{figure}

\clearpage
\begin{figure}
\begin{center}
\includegraphics[scale=1]{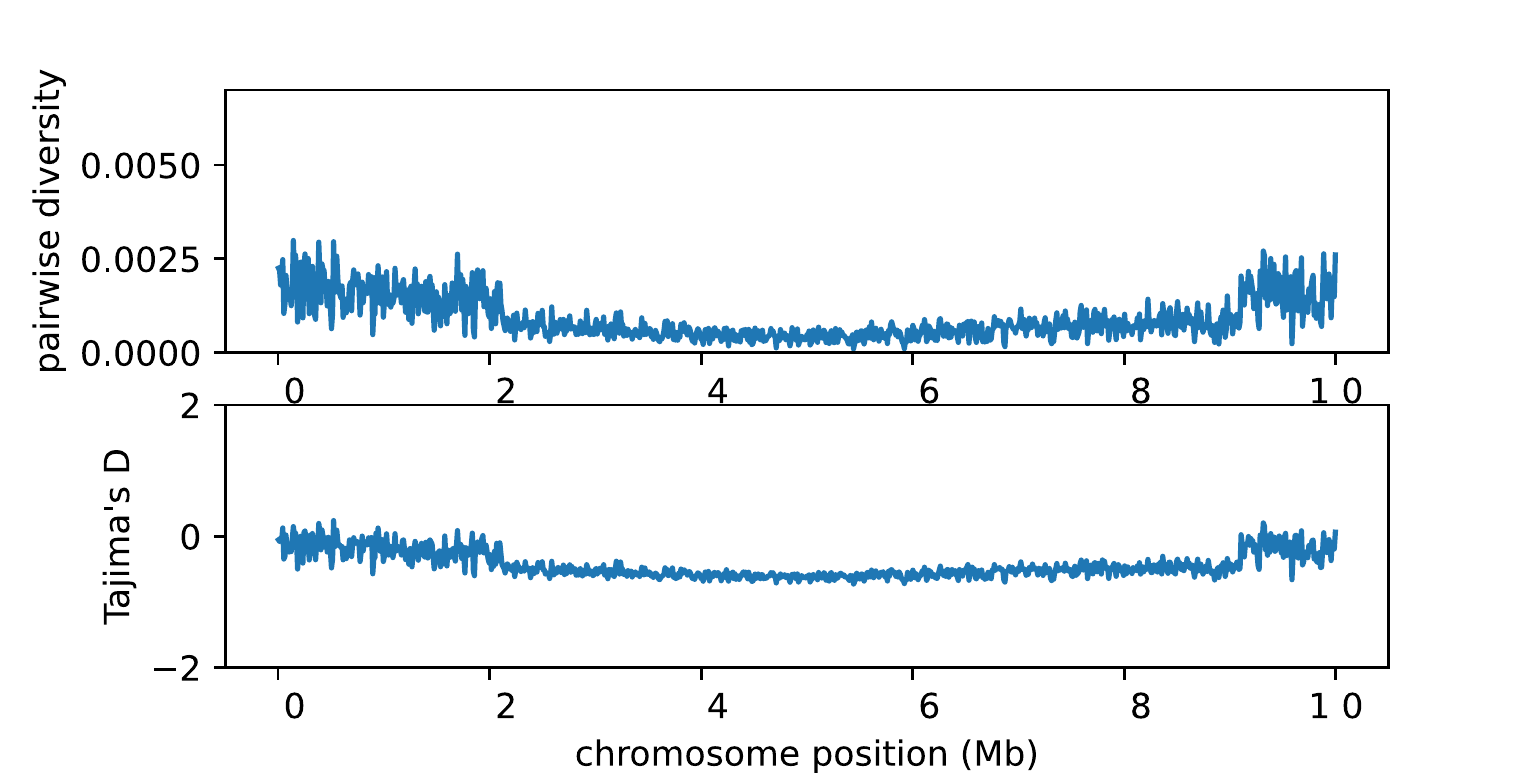}
\end{center}
\caption{Genetic diversity $\pi$ and Tajima's $D$ in response to a very strong selective sweep with $s=10$ after 20 generations under moderately reduced recombination. Simulations assume a population size of 300,000 individuals, with $r=5\times10^{-9}$. Under reduced recombination, the locus of the sweep does not move as close to zero diversity and background diversity is affected for the entire 10 Mb region and does not return to normal levels across the simulated contig, a pattern not observed in empirical data.  \label{SweepR5-9s10g120}}
\end{figure}
\clearpage

\clearpage

\clearpage

\begin{figure}
\begin{center}
\includegraphics[scale=.7]{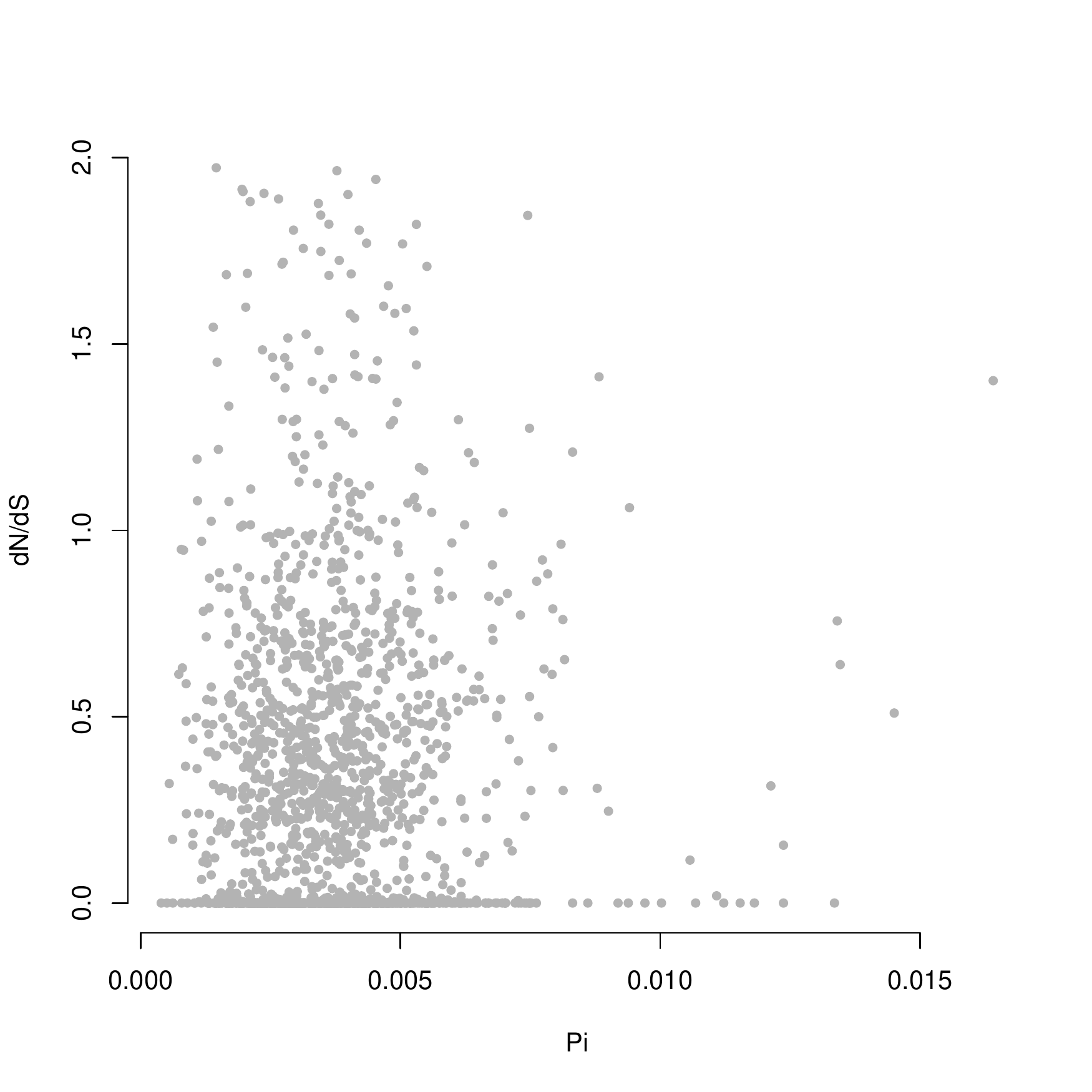}
\end{center}
\caption{Genetic diversity $\pi$ and dN/dS across paralogs are not correlated in \Mnerv, indicating that these two measures of selection can be applied independently in sequence data ($P=0.89$, $R^2=-0.00047$). Axes are truncated at dN/dS=2.0 for purposes of visualization. \label{PiDnDs}}
\end{figure}

%
%
%
%



\end{document}